\documentclass[
  aps,prx,
  reprint,
  superscriptaddress,
  nofootinbib,
  longbibliography
]{revtex4-2}

\usepackage{graphicx}
\usepackage{amsmath,amssymb,amsfonts}
\usepackage{booktabs}
\usepackage{bm}
\usepackage{xcolor}
\usepackage{xfrac}
\usepackage{hyperref}
\usepackage[nameinlink,noabbrev]{cleveref}

\crefname{equation}{Eq.}{Eqs.}
\Crefname{equation}{Equation}{Equations}
\crefname{section}{Sec.}{Secs.}
\Crefname{section}{Section}{Sections}
\crefname{appendix}{Appendix}{Appendices}
\Crefname{appendix}{Appendix}{Appendices}
\crefname{table}{Table}{Tables}
\Crefname{table}{Table}{Tables}
\crefname{figure}{Fig.}{Figs.}
\Crefname{figure}{Figure}{Figures}

\newcommand{\w}{\bm{\omega}}
\newcommand{\g}{\mathbf{g}}
\newcommand{\C}{\mathbf{C}}
\newcommand{\Cr}{\mathbf{C}_r}

\newcommand{\Lam}{\bm{\Lambda}}
\newcommand{\Id}{\mathbf{I}}
\newcommand{\etaag}{\eta_{\mathrm{agent}}}
\newcommand{\rhopair}{\rho_{\mathrm{pair}}}
\newcommand{\sread}{\sigma_{\mathrm{read}}}
\newcommand{\sidn}{\sigma_{\omega}}
\newcommand{\sseed}{\sigma_{u}}
\newcommand{\sign}{\mathrm{sign}}
\newcommand{\pibar}{\bar\pi}
\newcommand{\gain}{g}
\newcommand{\ract}{r_{\mathrm{act}}}
\newcommand{\rcomp}{r_{\mathrm{comp}}}
\newcommand{\Nag}{N_{\mathrm{a}}}
\newcommand{\rrf}{\bar r_{\mathrm{rf}}}
\definecolor{clsI}{HTML}{595959}
\definecolor{clsII}{HTML}{1F77B4}
\definecolor{clsIIIa}{HTML}{2CA02C}
\definecolor{clsIIIb}{HTML}{D62728}
\definecolor{clsIV}{HTML}{FF7F0E}
\definecolor{clsV}{HTML}{9467BD}
\newcommand{\swatch}[1]{\textcolor{#1}{\rule[0.28ex]{10pt}{1.4pt}}\;}
\newcommand{\smref}[1]{\cite[Sec.~#1]{SM}}
\newcommand{\smeq}[1]{\cite[Eq.~(#1)]{SM}}

\usepackage{enumitem}
\setlist[enumerate]{leftmargin=4ex, nosep, noitemsep}
\setlist[itemize]{leftmargin=4ex, nosep, noitemsep}

\begin{document}

\title{Role Differentiation as Ignition of a Collective Information Engine:\\  Structuration in Agent Populations}

\author{Maximilian Puelma Touzel}
\email{puelmatm@mila.quebec}
\affiliation{Mila--Quebec AI Institute, 6666 St.\ Urbain Ave., Montreal, Quebec, Canada}

\date{\today}

\begin{abstract}
Informational active matter shows how measurement-informed decisions produce collective order, so far in systems that reach consensus. We design collective information engines structured by differentiation instead, and construct a minimal instance using anti-coordination games where differentiated role information has value. Within many coexisting games, agents infer their role from a noisy social signal grounded in a persistent identity, and role-following action feeds back into that signal, which shapes the incentive to follow roles. Resources accrued through coordinated role-play combine with identity variability to reinforce the schemas that generated them. The model thereby operationalizes Sewell's duality of schemas and resources in Structuration, a resolution to structure--agency debates across social science. The engine ignites when a social loop gain---the product of identity persistence, cognitive capacity, channel fidelity, and schema strength---exceeds one. For a repertoire of such schemas, roles emerge with increasing gain in a bifurcation cascade whose functional form is fixed by the repertoire's eigenvalue spectrum, ranging from monitorable logarithmic sequences to avalanches that arrive without warning. Resource accumulation supplies the fitness of a replicator dynamics on schema strengths, which selects the cascade type endogenously. Subcritical identity covariance reveals that type before onset, enabling early detection, while feedback channel parameters bias which type is selected. Platform design then becomes a control lever to throttle emergent coordination. This theory grounds distributional AGI takeoff in a mechanism and provides a monitor-based solution. Joining game theory, collective dynamics, and information engines, we open a route to an information thermodynamics of agent populations.
\end{abstract}

\maketitle

\section*{Introduction}

\noindent A century-old question runs through the social sciences: how does macro-level social structure shape individuals while, simultaneously, individuals' choices generate that structure? Giddens proposed \textit{structuration} as a resolution to this \textit{structure--agency debate} based on the \textit{duality of structure} \cite{giddens1984}: social structure is both the medium and the outcome of the schemas and resources it organizes. Soon after, another sociologist, Sewell, noted that Giddens's duality accounted only for the reproduction of schemas---essentially a local stability condition rather than an endogenous dynamics. Motivated to provide a qualitative account of schema transformation, Sewell recast the duality as a feedback process between \emph{schemas}, the transposable rules and conventions that give meaning to resources by prescribing behavior around their use, and \emph{resources}, the material and symbolic stocks that schemas mobilize and that empower the agents using them \cite{sewell1992}.
Each sustains the other: schemas direct the use of resources, and resource-wielding agents reproduce (or transform) schemas. Debates resonant with the same micro--macro tension, many unresolved, recur across social science (\textit{e.g.} in cultural evolution \cite{acerbi2015} and in macroeconomics \cite{lucas1976,solow2010}). An explicit dynamics of structuration would help ground those debates in quantiative models.

Despite half a century of progress in agent modeling, Sewell's account of resource-schema structuration has no quantitative formalization. Existing agent modeling frameworks do not operationalize its core objects. Opinion dynamics has one state variable per agent and no resource that accumulates. Cultural evolution uses transmission and replicator models (\textit{e.g.} evolutionary game theory, including spatial reaction-diffusion extensions) that take the trait space or the game as given, without resource accumulation. Agent-based computational economics does accumulate resources under institutions \cite{tesfatsion2006}, but its institutions are rules imposed on the agents rather than objects the dynamics selects. More specialized models treat resource accumulation \cite{bouchaud2000} and payoff-dependent network reshaping \cite{santos2006}, but represent schemas as scalar state or payoff terms rather than as semantically loaded, transposable rules. None of these treats the resource-schema dynamics central to Sewell's proposal. Implementing the latter faces modeling decisions not specified by the qualitative account. The lack of an experimental system to empirically ground these choices lessens the scientific worth of any attempt.

Large Language Model-based multi-agent systems expand social simulation, putatively to both human social systems \cite{vezhnevets2023} and deployed AI agent systems \cite{anthropic2026}. They are also numerical systems suitable for precise experiments \cite{touzel2026position}. Calls to apply collective dynamics \cite{han2026} and systems theory \cite{miehling2025} now have responses: impressive applications of established frameworks, among them the statistical mechanics of opinion dynamics \cite{el2026} and replicator-like mean field theories \cite{flint2026}. Structuration, however, has no such framework to apply. Supplying one now would support research on a variety of timely questions about deployed AI, including agent personas and coordination.

We provide such a theory, exemplified for role differentiation in anti-coordination games, that introduces a new kind of informational active matter. Systems that convert measurement into extractable work through feedback \cite{parrondo2015,peliti2021}, subject to performance limits set by measurement quality \cite{Castro2026}, are known as \emph{information engines}. First demonstrated by Szilard almost 100 years ago \cite{szilard1929}, they are now understood as using information as a thermodynamic resource: measurement makes it available and feedback converts it into work \cite{barato2014,saha2023}. We show that a population of agents incentivized and provided with the means to role-play via $\omega$-identity signals can self-organize a collective, self-fueling information engine. Coordinated role-play is the payoff-generating output---the engine's analog of extracted work---that is bounded by the measured role information, just as Szilard's bound limits the work extracted by his illustrative model engine. As an instance of the long-noted link between Maxwell's demon and symmetry-breaking order parameters \cite{parrondo1999}, this description belongs to the emerging field of informational active matter: a modern effort to extend Szilard's ideas to more complex, in particular many-body settings, where measurement-informed local decisions produce collective order bounded by the information acquired \cite{vansaders2023, ziepke2022}. Those engines reach \emph{consensus}, a magnetization-like order. Each measurement here is of another agent, and informs which role to take in the interaction. The engine produces \emph{differentiation}, a covariance-like order in which agents take complementary rather than the same roles. The engine sustains social role structure by burning role information generated by that structure.

Modelling many-agent systems with tractable social structure is challenging. Schema transformation in the strict sense requires specifying a mechanism holding schemas on distinct axes. A breadth of choices differ in their assumptions and yield a family of theories (\cref{app:directions}). In this paper, we center the structuration that every such theory contains by scoping down onto the dynamics of schema strengths for a fixed repertoire to explain the schema--resource loop. Which schemas become institutionalized, in what order, and which agents are enriched are endogenous outcomes, while what a schema is \emph{about} is held fixed. Our main results are presented in dedicated sections:
\begin{itemize}
\item[I.]  a minimal model that couples schemas and resources through agent $\omega$-identities and actions;
\item[II.A.]  an ignition condition for self-sustaining role differentiation---an eigenvalue of the social loop gain $\Lam$ exceeding one, where $\Lam$ factorizes into $\omega$-identity persistence, agent cognitive capacity, social feedback channel fidelity, and the schema repertoire;
\item[II.B.]  a self-consistent stationary state, in which the loop closes into a condition on the schema repertoire alone, with resource weighting sustaining a diversity of schemas evoking complex society;
\item[III.]  a bifurcation cascade under extension to multiple schema contexts, in which role dimensions activate sequentially, with functional form determined by the eigenvalue spectrum of the schema repertoire;
\item[IV.]  an endogenous theory of that spectrum: a replicator dynamics on schema strengths, with fitness the resource-weighted variance and total regulated by a finite signaling capacity, which in the near-threshold mean field selects among cascade classes, from single avalanches to logarithmic cascades; and
\item[V.]  a monitoring protocol---subcritical spectral inversion of the $\omega$-identity covariance---that reads out the future cascade of the current repertoire before its onset, subject to finite-sample resolution and slow schema drift.
\end{itemize}
The first three results formulate a linearization about the symmetric fixed point that gives the loop gain $\Lam$, whose eigenvalues set the thresholds for role differentiation, and closed on itself it gives the stationary repertoire. \Cref{sec:cascade} classifies its spectrum. \Cref{sec:selects} decides which spectrum the loop produces. \Cref{sec:monitoring} recovers that spectrum from the subcritical covariance before the cascade begins. Full derivations are given in \crefrange{sec:gain}{sec:cascade} and in \cite{SM}.

We thus demonstrate that two ingredients suffice for roles to emerge: $\omega$-identities that persist and are legible to others---here lossy summaries of context-dependent action histories---and schema selection driven by the resource-weighted covariance of those $\omega$-identities. The role structure they produce is a collectivist solution to the coordination dilemma facing large agent populations\cite{jordan2025collectivist}: how to organize a role-based society that offers more payoff than the individually rational equilibrium (\cref{sec:model}). The discussion application to distributional AGI \cite{tomasev2026}: the hypothesis that general-level capability may emerge from a coordination phase transition among many individually-limited AI agents. The operator---the platform that runs the deployment and sets the channel and the repertoire---therefore decides before ignition which takeoff scenarios remain reachable. Finally, we set out what differentiation adds to informational active matter research. We argue that AI agent population systems are a promising experimental system in which structuration loops like the one articulated here can be built and measured in pursuit of developing a quantitative theory of agent populations.

\section{A minimal engine of social structure}\label{sec:model}

\begin{figure*}[t!]
\centering
\includegraphics[width=\textwidth]{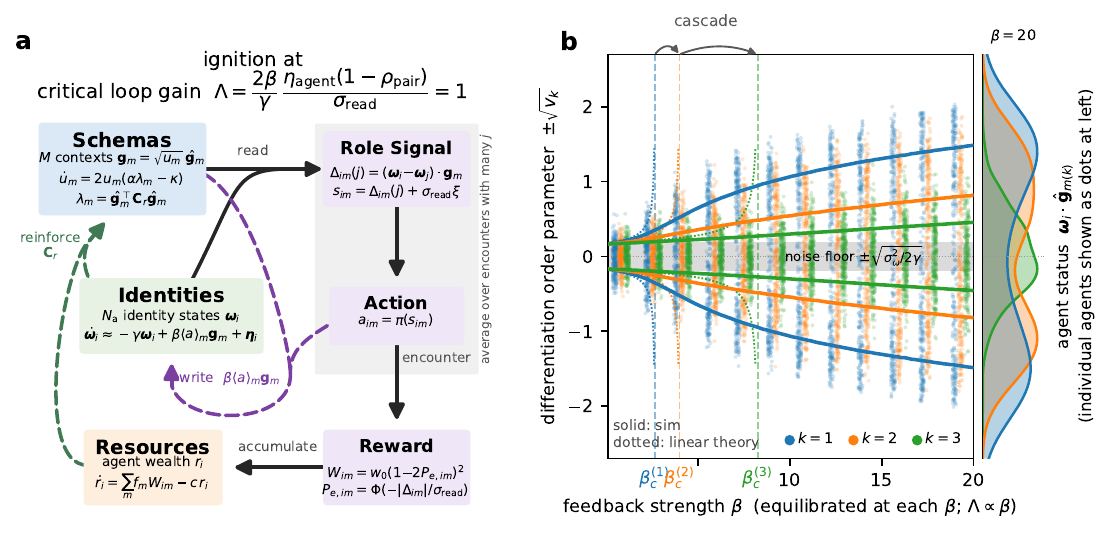}
\caption{\textbf{The self-consistent schema--role--resource loop and its ignition.}
\textbf{(a)} Model schematic, separating the three persistent state variables (schemas $\g_m=\sqrt{u_m}\,\hat\g_m$, of which only $u_m$ evolves; $\omega$-identities $\w_i$; resources $r_i$) from the fast per-encounter chain (status, action, reward). \emph{Solid} arrows are the feed-forward read-out: schema and $\omega$-identity form the status signal, the policy thresholds it into an action, and the coordination reward accumulates as resources. The shaded band is one encounter, averaged over the many partners $j$ drawn across encounters, after which the status difference enters only through $\Delta_{im}$. \emph{Dashed} arrows are the slow state updates that close the loop: resources reweight $\omega$-identities to reinforce schemas through the resource-weighted covariance $\Cr$, and actions are inscribed back into $\omega$-identities along the same axis (the write $\beta\langle a\rangle_m\g_m$). The single direction $\g_m$ is thus both read and write axis, which makes Sewell's schema--role duality a fixed-point loop, and the engine ignites when the loop gain $\Lambda$ of \cref{sec:gain} exceeds one (top).
\textbf{(b)} Three sequential bifurcations (differentiation order parameter vs.\ feedback strength) over the $\beta$ sweep ($\Nag=2500$, $d=M=3$, hierarchical fixed schema repertoire, random pairing; \smref{S10}). Points are individual agent statuses $\boldsymbol{\omega}_i\cdot\hat{\g}_{m(k)}$ along the three schema axes, with $m(k)$ the context whose strength gives the $k$-th eigenvalue. Solid curves are the order parameter $\pm\sqrt{v_k}$, the two branches of a pitchfork on each axis, with $v_k$ the $k$-th eigenvalue of the population covariance $\C$. Below its critical feedback strength an axis is unimodal within the noise band $\pm\sqrt{\sidn^2/2\gamma}$; above it (dashed) it splits into the complementary leader/follower role modes of the anti-coordination game, which separate further with $\beta$. Arrows on the top spine mark the cascade, the successive crossings of the critical feedback strengths $\beta_c^{(1)},\beta_c^{(2)},\beta_c^{(3)}$ labelled on the axis. Dotted curves are the linear mean-field theory, which diverges at $\Lambda_k=1$ (\cref{eq:cascade}), while the quartic confinement saturates the actual curve. The right strip shows the status distribution at $\beta=20$, axis 1 fully bimodal, axis 3 only marginally.}
\label{fig:loop}
\end{figure*}

\noindent \Cref{fig:loop}a schematizes the model described in this section to illustrate the closed loop: three slower varying state variables coupled through a fast per-encounter chain of status, action and reward. We build it in four steps: the game that makes roles worthwhile, the $\omega$-identities that realize them, the resources that accumulate, and the schemas that survive.

\subsection{The game}
\noindent We design utility into roles minimally, using a generic anti-coordination game. Two players choose to Go or Defer: both playing Go is costly to both ($-1,-1$), both deferring is mediocre ($3,3$), and complementary role-play pays best on average, though asymmetrically ($6,2$). The latter is played in a correlated equilibrium that pays $4$ per player (on average from random assignment), while the mixed Nash equilibrium pays only $2.5$. We realize this correlated equilibrium using a role signal as the correlating device. Even though role-following is not a dominant strategy---it is obeyed only when partners also follow---its incentive constraint is strict once roles differentiate and agents seek to maximize their payoffs (\smref{S1\,C}).
The maximal coordination gain $w_0 = 4-2.5 =1.5$ per encounter is the analogous work available to an engine that assigns complementary roles through the measurement of social signals. The specific numbers are inessential: any anti-coordination game enters the theory only through this single scalar $w_0$, which in turn sets merely the scale of wealth.
What is essential is the anti-coordination \emph{class}, in which roles are incentive-compatible: each action must be the best reply to its opposite---given Defer, Go must pay more ($6>3$), and given Go, Defer must pay more ($2>-1$). Even equal off-diagonal payoffs would suffice. We set them unequal to keep the general case. So what kind of engine, built from a multi-agent system, can assign complementary roles?

\subsection{Agents, game contexts, and $\omega$-identities}
\noindent $\Nag$ agents play the game against each other, selecting actions $a\in\{-1 (\mathrm{Defer}),+1 (\mathrm{Go})\}$ in pairwise encounters across a repertoire of $M$ distinct game contexts. We emphasize the importance of variable context in capturing both the \emph{transposability} of schemas and the context-dependent nature of roles. We represent $\omega$-identities, $\w_i$ for agent $i$, and contexts, $\g_m$ for context $m$, as vectors $\w_i,\g_m\in\mathbb{R}^d$ with embedding dimension $d$. Each agent stores and presents its own $\omega$-identity in each encounter. Two other storage architectures give the same encounter model: a platform-maintained \emph{status ledger}, and a \emph{platform estimate} reconstructed from the encounter log. They differ in which parameters an operator holds, and a platform estimate carries only the $\omega$-sector of the loop, not the resource and schema dynamics (\smref{S4}, \cref{app:control}). For simplicity, the $\omega$-identity of agent $i$ in context $m$ is contextualized as the scalar $\w_i\cdot\g_m$. When paired with $j$ in an encounter, agent $i$ makes a noisy observation of the \emph{status difference} $\Delta_{im}(j)=(\w_i-\w_j)\cdot\g_m$, through the signal $s_{im}=\Delta_{im}(j)+\sread\xi$ with $\xi$ a standard Gaussian noise source. We call this collective self-sensing process the \emph{status channel}. The channel serves to communicate context-dependent $\omega$-identity, relative to other agents in the context, making it useful as a correlating device of a correlated equilibrium, \emph{e.g.} prescribing Go ($a=+1$; payoff 6) where $\Delta_{im}(j)$ is positive and Defer ($a=-1$; payoff 2) where negative. The agent takes an action $a$ according to its action \emph{policy}, $\pi=p(a|\{s_{im,t}\}_{t=1}^T)$ after making a set of $T$ noisy status observations. The deterministic mapping $a=\sign(\tfrac1T\sum_{t}s_t)$ makes the Bayes-optimal, prescription-following decision under the uncertainty remaining after integrating these $T$ samples of evidence, with error rate depending on the magnitude of $\Delta_{im}(j)$ relative to $\sread$. The term \emph{status} reflects the reward asymmetry of the actions and the resulting $\omega$-identities. That asymmetry is immaterial for the theory, so the term should be seen generally as a mere label for the signal.

Roles require persistent $\omega$-identities. We realize this persistence simply by having each $\omega$-identity low-pass filter its agent's action history, here a context-dependent signed action input. This evolves under a restoring force and $\omega$-identity noise,
\begin{equation}
\dot{\w}_i = -\gamma\,\w_i(1+|\w_i|^2) + b_0\!\sum_{e\in E_i} a_e\,\g_{m_e}\,\delta(t-t_e) + \sidn\bm{\eta}_i ,
\label{eq:agent}
\end{equation}
where $E_i$ is the set of encounters involving agent $i$, $a_e$ and $m_e$ the decision and context of encounter $e$, $b_0$ the per-encounter kick, $\gamma$ the decay rate of $\omega$-identity, and $\bm{\eta}_i$ a vector of standard Gaussian noise sources. This noise represents influences on the $\omega$-identity outside the modeled encounters, and supplies the individual variation the loop amplifies into roles. \Cref{eq:agent} is an overdamped Langevin equation in a quartic confining potential, driven by encounter events: the restoring force is $-\nabla_{\w_i}\gamma(|\w_i|^2/2+|\w_i|^4/4)$, whose cubic term serves only to keep $|\w_i|$ finite (\cref{app:micro}). The cubic's coefficient is tied to $\gamma$ as a choice of units for $|\w_i|$, absorbed by rescaling $\g_m$ and $\beta$, so the linear decay rate is the only free parameter of the restoring force. Encounters are Poisson at rate $\nu$ per agent, a fixed interaction budget independent of system size $\Nag$, drawn across contexts with normalized frequencies $\sum_m f_m=1$ and over partners with pairing probabilities $p^{(m)}_{ij}$ normalized over the partner, $\sum_j p^{(m)}_{ij}=1$. Pairing is annealed: partners are redrawn at each encounter rather than fixed in a persistent network. A persistent network instead enters through $p^{(m)}_{ij}$ as a graph Laplacian (see \cref{app:micro} for this extension).

We derive a tractable version of \cref{eq:agent} through three approximations. The first coarse-grains the encounter stream over a window long compared with the average inter-encounter interval $1/\nu$ and short compared with the $\omega$-identity decay time $1/\gamma$. Such a window holds many encounters at a nearly fixed $\omega$-identity, each drawing its context, its partner, and its observation noise independently. Their empirical average therefore approaches the ensemble average over all three in the dense-encounter limit: $\nu\to\infty$, $b_0\to0$ at fixed $\beta\equiv b_0\nu$. Taking the limit discards shot noise, negligible where a single encounter moves an $\omega$-identity by much less than $\sidn^2/\beta$ (\cref{app:micro}). The window also contains the partner sum, so the drift no longer depends on which partners were drawn, only on the realized $\omega$-identities $\{\w_j\}$, through the status differences $\Delta_{im}(j)$. The second, the mean-field limit ($\Nag\to\infty$), replaces those realized values by the population distribution, for two reasons: the sum over $j$ concentrates on its mean, and the partner's $\omega$-identity loses the $O(1/\Nag)$ imprint of its past encounters with agent $i$. The third is weak $\omega$-identity noise. The steps so far leave the action averaged over the spread of partner statuses, and \cref{eq:agent-mf} collapses that spread to its mean. Subcritically the spread is $O(\sidn^2)$, so the collapse holds where it is small against $\sread^2$. Retaining it instead softens the gain by a factor known in closed form and shifts onset by $O(\sidn^2)$ (\cref{sec:gain}, \smref{S6}, \smref{S8}). The reduction has replaced $\Nag$ $\omega$-identity dynamics coupled through their encounters by one partner-independent dynamics, which each agent obeys separately. The dynamics becomes
\begin{equation}
\dot{\w}_i = -\gamma\,\w_i(1+|\w_i|^2) + \beta\sum_m f_m\,\langle a\rangle_m(\w_i)\,\g_m + \sidn\bm{\eta}_i ,
\label{eq:agent-mf}
\end{equation}
where $\langle a\rangle_m(\w_i)
%$ is the expected signed action in context $m$---a sigmoid of the true status difference, $\Delta_{im}$ (now averaged over the drawn partner $j$). The agent acts on the observed signal, $a=\pi(s_{im})$. The observation noise makes the action random given $\Delta_{im}$. Averaging the policy over that noise gives the Go-probability $\pibar(\Delta_{im})=\mathbb E_\xi[\pi(\Delta_{im}+\sread\xi)]=\Pr(a{=}{+}1\mid\Delta_{im})$, so $\langle a\rangle_m(\w_i)
=2\pibar(\Delta_{im})-1$ with $\Delta_{im}=\mathbb{E}_{\w_j}\left[(\w_i-\w_j) \cdot\g_m\right]$. Its slope at zero difference defines the \emph{read gain} $\gain\equiv2\pibar'(0)$, which sets how strongly actions are written into memory: thresholding the mean of $T$ Gaussian samples gives $\gain=\sqrt{2T/\pi}/\sread$.
The rate $\beta$ sets how strongly the social signal drives the $\omega$-identity, and is referred to below as the \emph{feedback strength}.

We assume the pairing probabilities $p^{(m)}_{ij}$ are set by a \emph{matching rule} that has access to $\omega$-identity information. The matching rule is unrestricted, and enters the linearized dynamics through one number, the correlation $\rhopair\in[-1,1]$ between a partner's status and one's own at fixed context $m$ (\cref{sec:gain}). Near the symmetric fixed point the pair is jointly Gaussian, so $\Delta_{im}=(1-\rhopair)\,\w_i\cdot\g_m$. Uniform matching draws the partner independently of agent $i$, so $\rhopair=0$ and the partner projection sits at its population mean, which vanishes in the role-free population. The coefficient is then one, and the partner contributes nothing. Rank-ordered (status-mirrored) matching drives $\rhopair\to-1$ and assortative (homophilic) matching drives it to $1$, the two ends of the range (\smref{S2\,B}). Both act on whatever status spread exists, so neither needs role structure to be defined.

\subsection{Resources}
\noindent Coordination income accumulates as agent wealth,
\begin{equation}
\dot r_i = \sum_m f_m W_{im} - c\,r_i, \;\;\textrm{with} \;\;W_{im}=w_0(1-2P_{e,im})^2
\label{eq:resource}
\end{equation}
the encounter-averaged payoff above Nash in context $m$ (\smeq{S2}), at role-inference error rate $P_{e,im}=\Phi(-|\Delta_{im}|/\sread)$ with $\Phi$ the standard Gaussian distribution function, and $c$ the resource-decay rate, so $1/c$ is the horizon over which past income still counts. This payoff obeys a Szilard bound, $W\le 2\ln2\,w_0\,I$ with $I=1-H_{\mathrm{bin}}(P_e)$ the role information supplied by the measurement (\smref{S1}). This is the differentiation counterpart of the information bound on consensus order in Ref.~\cite{vansaders2023}. This symmetric form of $I$ is the equiprobable two-role idealization. Heterogeneous status differences and pairwise error rates enter through $W_{im}$, the partner average of the pairwise payoff $W_{ijm}=w_0(1-2P_{e,ijm})^2$. The average runs over the same pairing statistics $p^{(m)}_{ij}$ that enter \cref{eq:agent}, with $W_{ijm}$ set by the magnitude of the status difference to the partner (extensions treated in \smref{S2}). In any case, $I$ bounds the coordination payoff, in information units. What a thermodynamic accounting would involve is set out in \smref{S9}.

\subsection{Schemas}
\noindent Agents who profit from taking differentiated roles along a schema reinforce it. A schema is a vector, on whose two parts reinforcement acts separately: its direction $\hat\g_m$ fixes what the schema is \emph{about}, while its magnitude fixes how strongly that prescription is \emph{institutionalized}. Writing $\g_m=\sqrt{u_m}\,\hat\g_m$ with $u_m=|\g_m|^2$ splits any such dynamics exactly into a \emph{strength sector} governing $u_m$---the reinforcement of a schema already in the repertoire---and a \emph{transformation sector} rotating $\hat\g_m$, the $\omega$-identity space direction along which anti-correlation in the paired agents' actions confers coordination benefit.

We develop the strength sector here, which suffices for schema-resource duality: it determines which of an available repertoire become institutionalized, in what order, and which agents are enriched. The schemas available to be played form the repertoire, and the environment fixes its context directions $\{\hat\g_m\}$. The environment also fixes where the schemas start: $u_m(0)=u_0e^{\sseed z_m}$, with $u_0$ the seed scale, $\sseed$ the log-dispersion across contexts, and $z_m$ standard Gaussian. For a deployed system $\sseed$ is set by the platform (\cref{app:control}). Restricting a dynamics to magnitudes along a fixed frame is standard practice---amplitude equations for a fixed set of critical modes \cite{cross1993}, replicator and Lotka--Volterra dynamics on a fixed set of types \cite{hofbauer1998}, the overlap coordinates of associative memories with quenched patterns \cite{krotov2016}, and the decoupled mode dynamics of deep linear networks under whitened inputs \cite{saxe2014} all take this form. Here the restriction is exact rather than approximate: \cref{eq:schema} is one half of a polar split of the vector dynamics. The other directional half, dynamics of the transformation sector, is analyzed in \cref{app:directions} where we show the dynamics and resulting solutions depend strongly on what additional assumptions are made about how schema directions interact, a question we consider separately.

A schema's strength $u_m$ acts at both stages of an encounter in its context. It scales the status difference read, $\Delta_{im}\propto\sqrt{u_m}$, which lowers the role-inference error rate $P_{e,im}$ against fixed noise $\sread$. It scales the kick written back into $\omega$-identity by the same factor, since that kick lies along $\g_m=\sqrt{u_m}\hat\g_m$. The two multiply, so the axis enters the loop gain of \cref{sec:gain} as $\gain u_m$. This fixes the feedback strength at which the status distribution in context $m$ bifurcates into Go and Defer role modes. Contexts therefore activate one at a time as the feedback strength rises, in the order of their strengths $u_m$. This sequence of bifurcations, the \emph{cascade} (\cref{sec:cascade}), is the focus of much of the rest of the paper.

The strengths also draw on one finite collective capacity: all $M$ schemas are read through the one status channel. Because the status difference read on axis $m$ scales as $\sqrt{u_m}$, the strength $u_m$ is the power that axis puts on the channel, and $U=\sum_m u_m$ is the total power the population uses for status signaling. A channel of finite power limits that total to $U_{\mathrm{cap}}$. We impose the limit softly, as a pressure $\varphi(U)$ common to every schema, negligible while the channel is slack and diverging as $U\to U_{\mathrm{cap}}$. Fixing $U$ instead would force the capacity to be exhausted and remove the slack regime in which schemas coexist below it (\cref{sec:closedpool}). Once the total reaches $U_{\mathrm{cap}}$, raising one strength lowers the rest: institutionalization is then the allocation of that capacity across the available games. In addition and independently, we impose that the channel has finite range: the status differences it can represent are bounded, and $\Delta_{im}$ grows as $\sqrt{u_m}$, so any single axis is capped, $u_m\le u_{\mathrm{cap}}$. Where power capacity $U_{\mathrm{cap}}$ is not saturated, schemas may reach their own range ceiling $u_{\mathrm{cap}}$ without displacing one another (\cref{sec:regimes}).

Each strength then grows in proportion to how much resource-weighted structure lies along its own axis, and decays at a rate common to every schema:
\begin{equation}
\begin{gathered}
\dot u_m/u_m = 2\bigl(\alpha\lambda_m-\kappa-\varphi(U)\bigr), \;\;\mathrm{with}\\
\lambda_m = \hat\g_m^{\!\top}\Cr\,\hat\g_m , \;\;
\g_m=\sqrt{u_m}\,\hat\g_m,\;\;\Cr = \frac{\sum_i r_i\, \w_i \w_i^{\!\top}}{\sum_i r_i}\\ \textrm{subject to} \;\; u_m\le u_{\mathrm{cap}}
\end{gathered}
\label{eq:schema}
\end{equation}
where $\Cr$ is the resource-weighted population covariance, $\lambda_m$ the resource-weighted variance along context $m$, $\kappa$ the context decay rate, $\alpha$ the context reinforcement rate, and $\varphi$ the budget pressure of the shared channel. Equivalently $d\ln u_m/dt=2(\alpha\lambda_m-\kappa-\varphi)$, the factor of two because $u_m=|\g_m|^2$ grows at twice the rate of $\g_m$.

The two ingredients of $\lambda_m$ encode Sewell's duality. It is \emph{covariance}-driven because a schema can organize only the variation that exists: only the differentiation already present along axis $m$ can be divided into complementary roles, so a schema is reinforced along the axes of differentiation the roles themselves produced. A schema aligned with no variance prescribes nothing and decays at rate $\kappa$. It is \emph{resource}-weighted because a schema is validated by the payoff its adherents accrue rather than by raw differentiation: weighting by wealth $r_i$ makes reinforcement contingent on success---the operational form of Sewell's claim that resources, accumulated by enacting a schema, sustain the schema that produced them. With reinforcement proportional to $u_m$, \cref{eq:schema} selects among existing schemas rather than creating new ones, at a per-unit rate set by $\lambda_m$. Because the decay is common to all schemas, the shares $u_m/U$ obey a replicator equation with fitness $\alpha\lambda_m$ exactly. The finite capacity regulates the total (\cref{sec:closedpool}). Transposability enters as the coherence of the repertoire, the overlaps $\hat\g_m\cdot\hat\g_n$, through which differentiation along one axis raises the resource-weighted variance along another (\cref{sec:closedpool}).

\Cref{eq:agent,eq:resource,eq:schema} together form the minimal social technology needed to close the perception-action-reward loop (\cref{fig:loop}a) in a way sufficient for social roles to emerge. The resulting multi-agent system dynamics formalizes
Sewell's vision and generates duality as a self-consistent fixed-point equation: schemas shape the role structure, roles generate resources, and resource-weighted structure ($\Cr$) selects schemas. The resource weighting makes survival contingent on success: a schema whose axis differentiates agents who profit by it will grow, while one whose adherents gain nothing decays at rate $\kappa$ however much $\omega$-identity varies along that axis (\cref{sec:fixedpoint}).
The generality of the dynamics suggests roles are not even a generic feature of the possible solutions. Next, we establish the transition across which they emerge.

\section{Ignition and the self-consistent state}

\noindent Here we solve the model in two subsections corresponding to two stages applied to a reduced state.
The full state is made from the $\omega$-identities $\{\w_i\}$ and resources $\{r_i\}$ over agents together with the strengths $\{u_m\}$ over contexts, evolving under \cref{eq:agent-mf,eq:resource,eq:schema}.
The $\w$-dynamics (\cref{eq:agent-mf}) is stochastic, so its stationary states are distributions, which we summarize by their moments. The $r$-dynamics (\cref{eq:resource}) has no noise, but inherits random variation from the $\omega$-identities through the status differences. The $u$-dynamics (\cref{eq:schema}) is deterministic, since its resource-weighted variance $\lambda_m$ is a population average.
The three dynamics run on separated timescales, $\gamma\gg c\gg\kappa$. $c\gg\kappa$ means the strengths see the resources at the stationary value $r^*$ set by the $\omega$-identities ($\w$-dynamics has no explicit dependence on $r_i$). The reduced state is then $(\w,\{u_m\})$, the $\omega$-identity coordinate of \cref{eq:agent-mf} together with the strengths.

We identify \emph{ignition} with the onset of instability around $\w=0$ that pushes each $\omega$-identity off the origin.
This onset can be determined by linearizing the dynamics of the reduced state. The $\w$-drift is odd, $\langle a\rangle_m$ being odd in $\Delta_{im}$, so the role-free density is symmetric under $\w\to-\w$ and its mean vanishes, $\langle\w\rangle=0$, at any values for the strengths. The $\omega$-identity noise spreads the population into a unimodal cloud about the origin (\cref{eq:C-linear}), whose status differences leave resources at the role-free mean $\rrf=O(\sidn^2)$ (\smref{S2\,B}).
Stationarity of the $u$-dynamics is easy to see from \cref{eq:schema}: either $u_m=0$ or $\alpha\lambda_m=\kappa+\varphi$, the \emph{invasion threshold} (\emph{i.e.} strength grows for $\alpha\lambda_m>\kappa+\varphi$). In the role-free phase $\varphi\approx0$ and that threshold is $\alpha\lambda_m=\kappa$. \Cref{sec:closedpool} treats the case where the budget binds.
The Jacobian in $(\w,\{u_m\})$ around $\w=0$ and at instantaneous strengths,
\begin{equation*}
\bm{J}=\begin{pmatrix}
-\gamma(\Id_{d\times d}-\Lam) & 0\\[2pt]
0 & 2\bigl[\alpha\,\mathrm{diag}(\bm\lambda)-\kappa\,\Id_{M\times M}\bigr]
\end{pmatrix},
\end{equation*}
is block-diagonal: the $d\times d$ $\omega$-identity block is the decay $-\gamma\Id_{d\times d}$ plus the linearized feedback matrix $\gamma\Lam$ of \cref{eq:linear}, where $\Lam$ is the \emph{social loop gain} (presented below);
the $M\times M$ strength block has $\bm\lambda$ that stacks the $\lambda_m$ of \cref{eq:schema} and is evaluated at the seed scale, where the cross-coupling term $2u_m\,\alpha\,\partial\lambda_m/\partial u_n$ is negligible against the diagonal. The elements of the two off-diagonal blocks are exactly zero. $\partial\dot\w/\partial u_m=0$ because the $\w$-dynamics depends on the strengths only through $\Lam\w$, which vanishes at $\w=0$. $\partial\dot u_m/\partial\w=0$ because $\Cr$ is even in $\w$, since $r^*$ and $\w\w^{\!\top}$ are both quadratic in it, so the strengths see the $\omega$-identity sector through the covariance alone and never through the population mean.
The $\omega$-identity block depends on the schema strengths only through $\Lam$, so the threshold criterion holds whether or not the strengths are stationary. We solve that block for given, fixed strengths in \cref{sec:gain}. Then, in \Cref{sec:fixedpoint}, we restore the schemas' own dynamics and solve \cref{eq:resource,eq:schema} for their stationary state. We state and summarize these results here before their derivation in subsequent subsections.

Being block diagonal, $\bm{J}$'s spectrum is the union of the spectra of the two blocks. Each provides a threshold expression for the onset of instability where one of its eigenvalues crosses zero: $-\gamma(1-\Lambda_k)$ at $\Lambda_k=1$ where the symmetric fixed point destabilizes along an eigenvector (the set of these crossings is the cascade \cref{eq:cascade}), and $2(\alpha\lambda_m-\kappa)$ at the invasion threshold, where the dynamics begins to reinforce a schema (\cref{sec:closedpool}).
Since the split into roles is symmetric, the population mean stays zero. We thus track the first threshold using the population covariance $\C=\Nag^{-1}\sum_i \w_i\w_i^{\!\top}$ set by $\bm{J}$ through the Lyapunov balance (\cref{eq:C-linear}).
We express $\Cr$ in terms of $\Lam$ and $\C$, and substitute for $\C$ the subcritical expression \cref{eq:C-linear} of \cref{sec:gain}, valid for $\Lambda_1<1$, so the loop closes below the first bifurcation.

The linearization gives the social loop gain
\begin{equation}
\Lam \equiv \underbrace{\frac{2\beta}{\gamma}}_{\substack{\text{identity}\\\text{persistence}}}\cdot
\underbrace{\etaag}_{\substack{\text{cognitive}\\\text{capacity}}}\cdot
\underbrace{\frac{1-\rhopair}{\sread}}_{\substack{\text{channel}\\\text{fidelity}}}
\underbrace{\bm{H}^{\!\top}\!\bm{H}}_{\text{repertoire}} ,
\label{eq:lambda}
\end{equation}
where $\bm{H}$ stacks the schema vectors $\g_m$, $\etaag\equiv\sread\gain/2$ is the agent's dimensionless inference efficiency, and $(1-\rhopair)$ is the pairing gain factor set by the matching rule's partner correlation of \cref{sec:model}. The first three factors are scalars, so they scale every eigenvalue of $\Lam$ alike and set the overall gain without distinguishing axes. Only the repertoire factor $\bm{H}^{\!\top}\bm{H}$ separates them: the symmetric fixed point destabilizes along eigenvector $k$ of $\Lam$ when the corresponding eigenvalue reaches one,
\begin{equation}
\Lambda_k > 1 ,\qquad \Lambda_1\ge\Lambda_2\ge\cdots\ge\Lambda_R ,
\label{eq:cascade}
\end{equation}
with $R=\mathrm{rank}(\bm{H})\le\min(M,d)$, short of $M$ when the pool outruns the $\omega$-identity dimension or the context directions are linearly dependent. Writing $\mu_1\ge\cdots\ge\mu_R>0$ for the eigenvalues of $\bm{H}^{\!\top}\bm{H}$, which are the schema strengths $u_k$ themselves when the contexts are orthogonal, $\Lambda_k$ is the leading gain $\Lambda_1$ scaled by $\mu_k/\mu_1$, so the ordered $\{\mu_k\}$ set the spacing of the thresholds, and $\Lambda_1$ sets how many have been crossed. The axes ignite one at a time where the eigenvalues are distinct, and together where the spectrum is degenerate. Here and in \cref{sec:cascade} the strengths are at their initial values, the environment's seeding of \cref{sec:model}, held fixed while $\beta$ sweeps. \Cref{sec:selects} makes them dynamical, so the spectrum then rises at fixed $\beta$ and crosses the same condition $\Lambda\mu_k=1$ from below, with $\Lambda$ the loop gain of a unit-strength axis (\cref{sec:gain}). $\Lambda_1=\Lambda\mu_1$ is the control parameter, a single dimensionless group collecting the write path $\beta/\gamma$, the read path $\etaag$, the channel factor $(1-\rhopair)/\sread$, and the leading repertoire eigenvalue $\mu_1$. The sweeps below vary $\beta$ as the accessible handle on that group. We make the population covariance $\C$ the order parameter.
Its eigenvalues $v_k$ leave the noise floor $\sidn^2/2\gamma$ one at a time. The number that have crossed, the height of the cascade $N$, is what we term \textit{thickness}: the count of active (supra noise floor) role dimensions. The \emph{thick} phase is where at least one axis has crossed, $N\ge1$.

For a single axis of unit strength $\Lam$ is a number and \cref{eq:cascade} is the loop-gain criterion of \cref{fig:loop}a, $\Lambda>1$. Ignition names the crossing of that criterion: above it the loop returns more differentiation than $\omega$-identity decay removes, so role structure sustains itself with no external drive. Here $\etaag$ is $1/\sqrt{2\pi}$ for a single observation and $\sqrt{T/2\pi}$ with memory $T$. $\etaag\to0$ is choice at random and $\etaag\to\infty$ deterministic payoff maximization. The inverse temperature spans the same two limits in the free-energy account of bounded rationality \cite{ortega2013}. The pairing enters the loop gain only through the factor $(1-\rhopair)$. Random matching is the reference, factor $1$. Rank-ordered matching gives factor $2$ in the population limit and halves the threshold, and assortative matching gives factor $0$ (\smref{S2}). The correlation needs no pre-existing role modes. $\rhopair$ is a property of the matching rule applied to whatever status distribution currently exists, including a unimodal and arbitrarily narrow one, so it is as well defined below threshold, where it sets the thresholds, as above it. Per-context play frequencies or matching reweight $\bm{H}^{\!\top}\bm{H}$ before its eigenvalues are taken, and then the channel factor no longer comes out in front (\smref{S2\,C}).

The engine runs on two integrators---a write path ($\beta/\gamma$: decisions accumulated into persistent $\omega$-identity) and a read path ($\etaag$: social signals accumulated into posterior inference)---coupled by the information channel between agents. The factors compensate at linear order: well-designed matching can compensate for limited-inference agents and vice versa (similar to the measurement-quality bound on engine performance derived in \cite{Castro2026}). However, even at the best (rank-ordered) matching ($\rhopair\to-1$) ignition requires $\etaag > \sread\gamma/4\beta$, an \emph{irreducible cognitive floor} below which no interaction orchestration can sustain role structure. At threshold where linear order is accurate, only the product $\Lambda$ is physical. Beyond this threshold the degeneracy of the factorization breaks: $\beta$ sets how far the roles separate, independently of $\rhopair$, while $\rhopair$ sets how often that separation converts to payoff. Role societies at equal $\Lambda$ but different splits are therefore dynamically distinguishable (\smref{S2}). \Cref{fig:loop}b shows the first three bifurcations in the cascade: equilibrating the population at each $\beta$, role axes leave the noise floor one at a time around the critical feedback strengths of \cref{eq:cascade}.

\subsection{The social loop gain and the cascade of thresholds}\label{sec:gain}

\noindent We linearize the mean-field drift \cref{eq:agent-mf} about the symmetric fixed point, showing how $\Lam$ arises. Its feedback term is $\beta\sum_m f_m\langle a\rangle_m(\w_i)\g_m$, with the expected signed action in context $m$ averaged over the observation noise and the drawn partner (\cref{sec:model}).

\paragraph{Mean field.} As $\Nag\to\infty$ the empirical distribution of the other agents converges to a fixed law, so the drawn partner is independent of the focal agent's own history. The matching rule fixes the partner correlation $\rhopair=\mathrm{Corr}(x_i,x_j)$ between the projections $x_i=\w_i\cdot\g_m$ and $x_j=\w_j\cdot\g_m$, and near the symmetric fixed point the pair is jointly Gaussian, so $\mathbb E[x_j\mid x_i]=\rhopair\,x_i$ and the expected status difference is $(1-\rhopair)x_i$ (\smref{S2\,B}). Random matching leaves the unconditional population mean, which vanishes at the symmetric fixed point, and is the case $\rhopair=0$. That difference is the $\Delta_{im}$ of \cref{sec:model}, $\Delta_{im}=(1-\rhopair)x_i$, so the observed signal is $s_{im}\approx\Delta_{im}+\sread\xi_{im}$, and, averaging the decision over the observation noise,
\begin{equation}\label{eq:abar}
\langle a\rangle_m(\w_i)=2\pibar(\Delta_{im})-1,
\end{equation}
which for the single-sample policy is $\operatorname{erf}\!\bigl(\Delta_{im}/(\sread\sqrt2)\bigr)$. Substituting the conditional mean inside the policy is exact as $\sidn\to0$, where the cloud shrinks to a point at the origin and the threshold is set; the correction away from onset is $O(\sidn^2)$ (\smref{S6}).

\paragraph{Linearization.} At $\w_i=\bm0$ each projection vanishes and $2\pibar(0)-1=0$ by policy symmetry. To first order $\langle a\rangle_m(\w_i)\approx \gain(1-\rhopair)(\g_m^{\!\top}\w_i)$, so the feedback term of \cref{eq:agent-mf} becomes
\begin{equation}
\begin{split}
\beta\sum_m f_m\langle a\rangle_m(\w_i)\g_m&\approx \beta\gain(1-\rhopair)\Bigl(\sum_m f_m\,\g_m\g_m^{\!\top}\Bigr)\w_i\\
&=\gamma\,\Lam\,\w_i,
\end{split}
\end{equation}
\begin{equation}
\begin{gathered}
\Lam\equiv \Lambda\sum_m f_m\,\g_m\g_m^{\!\top}=\Lambda\,\bm{H}^{\!\top}\mathrm{diag}(\mathbf{f})\,\bm{H},\\[2pt]
\Lambda\equiv\beta(1-\rhopair)\gain/\gamma,
\end{gathered}
\end{equation}
with $\mathbf{f}=(f_1,\ldots,f_M)$, $\bm{H}$ the $M\times d$ matrix of stacked context vectors, and $\Lambda$ the scalar loop gain of a unit-strength axis. We take the play frequencies uniform, $f_m=1/M$, and absorb the constant $1/M$ into $\beta$, so that $\Lam$ takes the form of \cref{eq:lambda} and $\beta$ is the per-context write rate. The frequencies are retained explicitly only where they act as a design lever (\smref{S2\,D}). They are properties of the environment, fixed over the full set of $M$ candidate contexts, and are not renormalized when a schema's strength decays away. Its context is still played, so extinction does not concentrate weight on the survivors (\cref{sec:closedpool}). Combined with the linearized restoring force (the cubic term vanishes at the origin),
\begin{equation}\label{eq:linear}
\dot{\w}_i\approx-\gamma\,(\Id_{d\times d}-\Lam)\,\w_i+\sidn\bm\eta_i.
\end{equation}
The matching rule enters only through the factor $(1-\rhopair)$, which multiplies every axis alike and vanishes at $\rhopair=1$. The symmetric fixed point thus loses stability along eigenvector $k$ of $\Lam$ when its eigenvalue reaches one (\cref{eq:cascade}), so the thresholds are
\begin{equation}\label{eq:thresholds}
\Lambda_k=\Lambda_1\frac{\mu_k}{\mu_1}=1,\qquad \mu_1\ge\mu_2\ge\cdots\ge\mu_d,
\end{equation}
the cascade of thresholds, crossed in the order of the repertoire's eigenvalues.

\paragraph{The subcritical state.} At fixed schemas the drift is the gradient $-\nabla_{\w}E$, with
\begin{equation}\label{eq:energy}
\begin{gathered}
E(\w)=\gamma\Bigl(\tfrac{|\w|^2}{2}+\tfrac{|\w|^4}{4}\Bigr)-\beta\sum_m\Psi(\w\cdot\g_m),\\[2pt]
\Psi'(z)=2\pibar\bigl((1-\rhopair)z\bigr)-1,
\end{gathered}
\end{equation}
so the stationary distribution is Boltzmann, $p^*(\w)\propto\exp(-2E(\w)/\sidn^2)$, and $\omega$-identities are i.i.d.\ draws. Dropping the quartic term recovers the linear drift of \cref{eq:linear}, so near threshold $p^*$ is Gaussian with covariance (Lyapunov equation, symmetric drift)
\begin{equation}\label{eq:C-linear}
\C=\frac{\sidn^2}{2\gamma}(\Id_{d\times d}-\Lam)^{-1},
\end{equation}
valid for $\Lambda_1<1$. $\C$ then has the same eigenbasis as $\Lam$ and thus as $\bm{H}^{\!\top}\bm{H}$, with eigenvalues
\begin{equation}\label{eq:vk}
v_k=\frac{\sidn^2}{2\gamma(1-\Lambda_k)},
\end{equation}
so $\C$ diverges along the leading eigenvector of $\Lam$ at the bifurcation. Each $v_k$ increases with the loop gain $\Lambda_k$ it comes from, so \cref{eq:vk} can be inverted: measuring the covariance spectrum below threshold recovers $\{\Lambda_k\}$, and with it the cascade still to come (\cref{sec:monitoring}).

\emph{Validity of the linearization.} Stability is a Jacobian property evaluated \emph{at} the fixed point, independent of where the probability mass sits, so \cref{eq:thresholds} is the exact stability boundary at any noise level. The cloud's width enters only through the partner average, which softens the gain and shifts onset by $O(\sidn^2)$. This is a mean-field Ginzburg correction that vanishes with $\sidn$ and is negligible outside a shrinking critical window (\smref{S6}).

% ======================================================================

\subsection{The self-consistent fixed point}\label{sec:fixedpoint}
% ======================================================================

\noindent \Cref{sec:gain} solved the $\omega$-identity sector. The two slower dynamics close the loop. \Cref{eq:agent-mf,eq:resource,eq:schema} evolve on separated timescales $\gamma\gg c\gg\kappa$, so $\omega$-identities equilibrate first, then resources, then schemas, and each is stationary with respect to the ones below it.

\paragraph{Step 1: resources.} At steady state $r_i^*=c^{-1}\sum_m f_m\,\mathbb{E}_{j}[W_{ijm}]$, a deterministic function of $\w_i$. Near threshold the error rate linearizes, $P_e\approx\tfrac12-|(\w-\w')\cdot\g_m|\pibar'(0)$, so $W_{ijm}\approx w_0\gain^2((\w_i-\w_j)\cdot\g_m)^2$. Averaging over the drawn partner, jointly Gaussian with the focal projection at correlation $\rhopair$ ($\mathbb{E}[x_j\mid x_i]=\rhopair x_i$ and $\mathrm{Var}[x_j\mid x_i]=(1-\rhopair^2)\,\g_m^{\!\top}\C\g_m$, with $x_i=\w_i\cdot\g_m$),
\begin{equation}\label{eq:rstar}
\begin{split}
r^*(\w)\approx\frac{w_0\gain^2}{c}\sum_m f_m\Bigl(&(1-\rhopair)^2(\w\cdot\g_m)^2\\
&+(1-\rhopair^2)\,\g_m^{\!\top}\C\g_m\Bigr).
\end{split}
\end{equation}
The first term is agent-specific; the second a population baseline. Random matching removes both factors.

\paragraph{Step 2: resource-weighted covariance.} With $\Cr=\int r^*(\w)\w\w^{\!\top}p^*/\int r^*(\w)p^*$ and $r^*$ quadratic in $\w$, the numerator is a fourth moment, evaluated for Gaussian $p^*$ by Wick's theorem (Isserlis' theorem in the statistics literature),
\begin{equation}\label{eq:isserlis}
\mathbb{E}[(\w\cdot\g_m)^2\,\omega_a\omega_b]=(\g_m^{\!\top}\C\g_m)\,C_{ab}+2(\g_m^{\!\top}\C)_a(\g_m^{\!\top}\C)_b,
\end{equation}
making $\Cr$ a function of $\C$ and $\{\g_m\}$ alone. Even when $\C\propto\Id_{d\times d}$, $\Cr$ is anisotropic, because $r^*(\w)\propto\sum_m f_m(\w\cdot\g_m)^2$ weights agents whose $\omega$-identities align with the context directions more heavily. This anisotropy resolves the collapse of the unweighted dynamics ($r_i\equiv1$, $\Cr=\C$): $\Cr$ inherits structure from the repertoire even when the population is isotropic. Summing \cref{eq:isserlis} over contexts gives a closed form for the resource-weighted covariance, valid for any repertoire geometry and unchanged when $\Lam$ is rescaled,
\begin{equation}\label{eq:Cr-closed}
\Cr=\C+(1-\rhopair)\,\frac{\C\,\Lam\,\C}{\operatorname{tr}(\C\Lam)}.
\end{equation}
$\C$ and $\Lam$ share an eigenbasis (\cref{eq:vk}), so $\Cr$ has the same one as $\bm{H}^{\!\top}\bm{H}$. The two prefactors of \cref{eq:rstar} combine into the single factor $(1-\rhopair)$ (\smref{S2\,B}), so assortative matching ($\rhopair\to1$) leaves $\Cr=\C$ and removes the resource weighting altogether, while rank-ordered matching doubles it. For orthogonal contexts \cref{eq:Cr-closed} collapses to a scalar relation on each axis,
\begin{align}\label{eq:lambda-m}
\lambda_m=v_m\Bigl(1+(1-\rhopair)\frac{u_mv_m}{\sum_n u_nv_n}\Bigr),\\
\mathrm{with}\;\; v_m=\frac{\sidn^2}{2\gamma(1-\Lambda_m)},\nonumber
\end{align}
where $v_m=\hat\g_m^{\!\top}\C\hat\g_m$ is the eigenvalue of $\C$ along $\hat\g_m$, the $v_k$ of \cref{eq:vk} indexed by context rather than by rank. The strength enters through $\Lambda_m=\Lambda u_m$. The resource weighting fixes that denominator: normalizing $\Cr$ by $\sum_ir_i$ puts the population's mean income there, so $\sum_nu_nv_n$ is the role-free mean $\rrf$ up to constants (\smref{S2\,B}).

\paragraph{Step 3: schemas.} At steady state each surviving schema satisfies $\alpha\lambda_m=\kappa$, with $\lambda_m=\hat\g_m^{\!\top}\Cr\hat\g_m$ the resource-weighted variance along its (fixed) axis. Those with $\lambda_m<\kappa/\alpha$ decay to zero. Steps 1--3 therefore close into a condition on the repertoire alone,
\begin{equation}\label{eq:selfcon}
\alpha\,\lambda_m\bigl(\{\g_n\}\bigr)=\kappa \qquad \text{for every surviving } m,
\end{equation}
since \cref{eq:C-linear,eq:Cr-closed} fix $\Cr$ from $\{\g_n\}$ and nothing else. Schemas set the $\omega$-identities whose resource-weighted covariance selects the schemas, which is Sewell's duality as a fixed-point equation. For orthogonal contexts \cref{eq:lambda-m} makes it explicit in the strengths $\{u_n\}$.

\paragraph{The loop and multiplicity.} The conditions close: $\{\g_m\}\!\to\! E\!\to\! p^*\!\to\!\C\!\to\! r^*\!\to\!\Cr\!\to\!\{\g_m\}$. The trivial solution $\g_m^*=0$ always exists. Nontrivial solutions appear when an eigenvalue of $\Cr$ exceeds $\kappa/\alpha$. Each is a self-consistent fixed point, a solution of \cref{eq:selfcon}. Because $\Cr$ is anisotropic even at isotropic $\C$, those eigenvalues differ from the outset, so the condition selects among schemas rather than admitting or rejecting all of them. Since $\Cr$ has $d$ eigenvalues, the sustained schema count is
\begin{equation}\label{eq:Meff}
M_{\mathrm{eff}}=\#\{k:\lambda_k(\Cr)\ge\kappa/\alpha\}\le d,
\end{equation}
at such a fixed point the number of axes whose resource-weighted variance clears the context decay rate. The bound counts eigendirections, so it holds while distinct roles occupy distinct axes of $\Cr$. Where schemas may overlap, the ceiling is instead the associative-memory capacity $d^{\,n-1}$ of \smref{S3}, which exceeds $d$ because stored patterns need not be orthogonal. It caps the cascade: $N$ cannot exceed the number of axes available to activate, and reaches $M_{\mathrm{eff}}$ as $\beta$ grows. How many of them the budget can sustain, and whether the endpoint is a saturated cluster ($N=M_{\mathrm{eff}}$) or a single condensate ($N=1$), is decided by the capacity ratio of \cref{sec:closedpool}. Without resources ($r_i\equiv1$, $\Cr=\C$) the $\lambda_m$ rank the fixed axes by their alignment with the leading eigenvector of $\C$, so the survivor set (\cref{sec:closedpool}) narrows toward the best-aligned schema. The resource reweighting, anisotropic along the context directions themselves, lifts several $\lambda_m$ over $\kappa/\alpha$ at once and sustains a multi-schema state. This fixed point identifies the surviving schemas but leaves the shape of their spectrum implicit. Since this shape determines the cascade, \Cref{sec:cascade} asks what an arbitrary spectrum implies for the cascade, and classifies the possible cascades. \Cref{sec:selects} analyzes the capacity-constrained dynamics and determines which spectra and cascade result. The fixed point organizes the dynamics rather than being a solution they reach. \Cref{sec:closedpool} shows it is a saddle, which the solved trajectories leave along the unstable direction.

% ======================================================================

\section{The functional form of the cascade}\label{sec:cascade}

\begin{figure*}[t!]
\centering
\includegraphics[width=\textwidth]{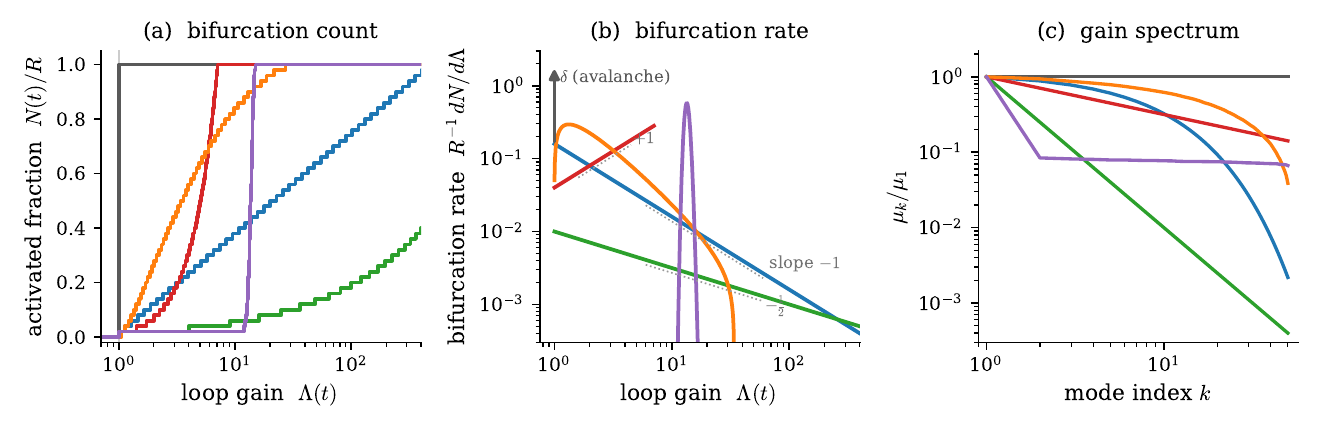}\\[8pt]
\makebox[\textwidth][l]{\textbf{(d)}}\\[2pt]
\begin{tabular}{@{}llll p{4.6cm} l@{}}
\toprule
Cascade class & $N(\beta)$ & Spectrum & $\mu_k$ & Repertoire & For a monitor \\
\midrule
\swatch{clsI}(i) single avalanche & step & degenerate & $\mu_k=\mu$ & \raggedright orthogonal, flat strengths & no warning \\
\swatch{clsII}(ii) logarithmic & $k_0\ln\Lambda$ & geometric & $\mu_k\propto e^{-k/k_0}$ & \raggedright orthogonal, exponential strengths & steady per $e$-fold \\
\swatch{clsIIIa}(iii) decelerating & $\Lambda^{1/a}$ & Zipf $a>1$ & $\mu_k\propto k^{-a}$ & \raggedright orthogonal, power-law strengths & informative early \\
\swatch{clsIIIb}(iii) accelerating & $\Lambda^{1/a}$ & Zipf $a<1$ & $\mu_k\propto k^{-a}$ & \raggedright orthogonal, power-law strengths & early rate too low \\
\swatch{clsIV}(iv) quiet, then burst & $(\Lambda-1)^{3/2}$ & random & Mar\v{c}enko--Pastur & \raggedright random directions, aspect ratio~$q$ & loop still open \\
\swatch{clsV}(v) one, gap, then all & gap, then jump & spiked & outliers $+$ bulk & \raggedright orthogonal, one dominant strength; or coherence $\chi$ & quiet mimics an end \\
\bottomrule
\end{tabular}
\caption{\textbf{Cascade classes of exogenous schema repertoires.} \textbf{(a)} The number of active role axes $N$ as the loop gain increases in time, $\Lambda(t)$. \textbf{(b)} The bifurcation rate over the same sweep, which accelerates where its slope is positive, equivalently where $\zeta>2$ (\cref{eq:accel}). \textbf{(c)} The rank-ordered gain spectrum, normalized to its leading eigenvalue. Panel (a) counts how many of them have bifurcated at each gain. \textbf{(d)} The classes and their cascades, the last column anticipating the monitoring analysis of \cref{sec:monitoring}. All four are given over the five spectral classes of the repertoire $\bm{H}^{\!\top}\bm{H}$, keyed by the colours of (d): (i) degenerate spectra give a single avalanche with no advance warning; (ii) geometric hierarchies give a logarithmic cascade with constant bifurcations in log-space; (iii) power-law (Zipf) hierarchies split at exponent $a=1$ into decelerating (informative early phase) and accelerating (early rate under-predicts) cascades, with log--log slope $(1-a)/a$ in panel (b); (iv) unstructured random repertoires (Mar\v{c}enko--Pastur) start quietly, $dN/d\Lambda\propto(\Lambda-1)^{1/2}$, then burst; (v) coherent (spiked) repertoires produce an early bifurcation, a quiescent interval, then an avalanche. The curves use $k_0=8$, $a=2$ and $a=\tfrac12$, $q=2$, and $\chi=0.2$, all drawn with the same $R=50$ modes, the random class as $50$ quantiles of its bulk, so only the spectral shape differs. The random and spiked classes are single realizations at a fixed seed. The quiescent interval reads out the gap below the leading mode in (c), and the crowding of the random spectrum against its upper edge makes the early rate vanish.}
\label{fig:classes}
\end{figure*}

\noindent We consider sweeps of increasing loop gain, $\Lambda(t)$, over time, loosely summarizing the culture and technology innovations that improve $\omega$-identity persistence, inferential capabilities of individuals, and/or social channel fidelity leading to role-based society.
In our theory, only the product is physical at threshold, so raising any factor of $\Lambda$ serves equally to drive the emergence of roles and structured use of a repertoire of schemas. For our simulations, we sweep $\Lambda$ by sweeping $\beta$, which enters $\Lambda$ linearly. We assume the onset of role-based society is slow relative to the set of $\omega$-identities taken on by individuals in a population, and so take the sweep to be slow compared with $\omega$-identity relaxation, $\dot\Lambda/\Lambda\ll\gamma$, so the population tracks it. What follows are the cascades possible for any such fixed spectrum. \Cref{sec:selects} asks which spectra the loop produces.

The number of active axes at feedback strength $\beta$ is the counting function $N(\beta)$, whose local rate of growth follows from the same construction. Both are fixed by the eigenvalue spectrum $\rho$ of the repertoire $\bm{H}^{\!\top}\bm{H}$ in the limit $M,d\to\infty$ at fixed aspect ratio $M/d$. Let $\mu_1\ge\cdots\ge\mu_R>0$ be its nonzero eigenvalues, $R=\mathrm{rank}(\bm{H})\le\min(M,d)$, so that $\Lambda_k=\Lambda\mu_k$ with $\Lambda$ the scalar prefactor of \cref{eq:lambda}. Axis $k$ bifurcates at $\Lambda=1/\mu_k$ (\cref{eq:cascade}), so at loop gain $\Lambda$
\begin{equation}\label{eq:counting}
N(\Lambda)=\#\{k:\mu_k>1/\Lambda\}=R\,\bar F(1/\Lambda),
\end{equation}
\begin{equation}
\bar F(\mu)=\int_{\mu}^\infty\rho(\mu')\,d\mu',
\end{equation}
with $\rho$ the spectral density of $\bm{H}^{\!\top}\bm{H}$, so $N$ counts the eigenvalues the gain has passed. The rate is
\begin{equation}\label{eq:rate}
\frac{dN}{d\Lambda}=\frac{R}{\Lambda^{2}}\rho\!\Bigl(\frac{1}{\Lambda}\Bigr),\qquad
\frac{dN}{d\ln\Lambda}=R\,\mu\rho(\mu)\big|_{\mu=1/\Lambda}.
\end{equation}
Defining the local spectral exponent $\zeta(\mu)\equiv-d\ln\rho/d\ln\mu$, differentiating again gives the acceleration criterion
\begin{equation}\label{eq:accel}
\frac{d^2N}{d\Lambda^2}>0\iff\frac{d}{d\mu}[\mu^2\rho(\mu)]<0\iff \zeta>2 \ \text{at}\ \mu=1/\Lambda.
\end{equation}
The criterion \cref{eq:accel} is local, since $\zeta$ is evaluated at $\mu=1/\Lambda$, the point the sweep has reached. A cascade with $\zeta<2$ there can still accelerate further down the spectrum. This limits what $\zeta$ tells a monitor about the sequence still to come (\cref{sec:monitoring}).

The three properties of the repertoire shape $\rho$: the strength hierarchy $\{u_m\}$, the coherence $\{\hat\g_m\cdot\hat\g_n\}$, and the aspect ratio $q=M/d$. \Cref{eq:schema} makes the first endogenous and leaves the other two to the environment. Each of the five classes catalogued in \cref{fig:classes}d is the canonical form taken when one of the three dominates, and the cascade induced by the canonical form follows from \cref{eq:counting}. Any repertoire's spectrum is locally one of these, and exactly one where the spectrum is quasistatic across the sweep. \Cref{fig:classes}a,c give an example of each at $R=50$ modes, while \cref{fig:classes}b shows the rate for $R\to\infty$.

The boundary $a=1$ ($\zeta=2$), with $a$ the Zipf exponent of the strength hierarchy, separates the decelerating and accelerating cascades of a Zipf-shaped spectrum. The random class \cite{marcenko1967} has its spectrum confined to $[\mu_-,\mu_+]$, with the edges set by $q$. Bifurcations run from the largest eigenvalue down, so the cascade enters the bulk at its upper edge, where the density vanishes as $\rho\sim\sqrt{\mu_+-\mu}$---hence the vanishing onset rate $dN/d\Lambda\propto(\Lambda-1)^{1/2}$ before the bulk. The spiked class comes from redundancy rather than dominance: coherent contexts sharing directions. At uniform coherence $\chi$ and common strength $u=|\g_m|^2$, $\bm{H}^{\!\top}\bm{H}$ has one outlier eigenvalue $u(1+(M-1)\chi)$ above a degenerate bulk at $u(1-\chi)$, and the gain scales both by $\gain$. The $\Lambda(t)$ sweep crosses the outlier first and the bulk all at once a factor $(1+(M-1)\chi)/(1-\chi)$ later, so a quiescent interval of that size separates the first bifurcation from a step in $N$.

\section{Endogenous selection of the repertoire}\label{sec:selects}

\noindent In \cref{sec:cascade} the schema strengths, and thus the cascade type, were taken as given. $\{u_m\}$ are set instead by the slower schema-resource dynamics of \cref{eq:resource,eq:schema}, both at the stationary state and along the $u$-transient that approaches it. The context directions $\{\hat\g_m\}$ stay exogenous throughout (\cref{sec:model}). Two ratios separate the regimes, one from the $u$-transient and one from the stationary state. The first, $r=\ract/\rcomp$, is fixed by the dispersion of the seeded repertoire and is defined with the transient in \cref{sec:regimes}. The second is the \emph{capacity ratio} $\Theta=(U_{\mathrm{cap}}/M)/u_{\mathrm{cap}}=M_{\mathrm{cap}}/M$, which measures the budget's equal share per schema against the per-schema ceiling, with $M_{\mathrm{cap}}=U_{\mathrm{cap}}/u_{\mathrm{cap}}=\Theta M$ the number of schemas the budget sustains at that ceiling. \Cref{sec:closedpool} finds the stationary states of the closed pool and shows that the symmetric one is a saddle that funnels approaching trajectories away (\cref{fig:saddle}). \Cref{sec:regimes} then places those endpoints on the parameter space slice spanned by those two ratios and reads off the cascade each produces to obtain a phase diagram (\cref{fig:regimes}).

% ======================================================================

\subsection{The closed-pool (survivor) construction}\label{sec:closedpool}

\noindent Under the full loop the strength dynamics produce the spectrum. We treat the $M$ context directions as candidate schemas and ask how many survive. The pool is \emph{closed}: no schema is added after seeding, so the dynamics can only prune it. Schema innovation is an open-repertoire process we do not treat (\cref{sec:regimes}).

\paragraph{The active set.} The active set is $\{m:u_m>u_0\}$ with $u_m=|\g_m|^2$ and $u_0$ the seed scale, the small strength at which schemas are seeded and above which they count as active. A schema \emph{survives} when its strength stays finite at the end of the \emph{$u$-transient}, which institutionalizes the repertoire by winnowing the seeded pool to its survivors and setting the shape of their strength spectrum (\cref{sec:regimes}). The set's size $K\le M$ is the emergent survivor count. Orthogonal contexts make each $\hat\g_m$ an eigenvector of $\Cr$ (\cref{sec:fixedpoint}), so $\lambda_m$ is an eigenvalue of $\Cr$. Fixed axes keep distinct survivors on distinct eigen-directions: a schema cannot rotate onto another, so two survivors share a direction only if the repertoire supplies them one. In this case, $K$ equals $M_{\mathrm{eff}}$ (\cref{eq:Meff}), the number of eigenvalues of $\Cr$ that clear $\kappa/\alpha$. Under the vector dynamics of \cref{app:directions} $K$ and $M_{\mathrm{eff}}$ would instead separate, since every schema rotates toward the same leading eigenvector. The seeding of \cref{sec:model} is centered on $u_0$, so half the repertoire starts above it at any $\sseed$. Only the $u$-transient separates the $u_m$ into a saturated group and a group decaying to zero, and the count is insensitive to $u_0$ once it has. But $u_0$ does enter the regime boundaries through the log dynamic range $L=\ln(u_{\mathrm{cap}}/u_0)$ of \cref{sec:regimes}. The cascade height never exceeds the survivor count, $N\le K$, because a schema survives at $u_m>u_0$ but ignites only at $\Lambda u_m>1$. The two coincide once the transient has saturated the survivors at $u_{\mathrm{cap}}$, provided $\Lambda u_{\mathrm{cap}}>1$.

\paragraph{The strength replicator.} In log form the strengths of \cref{eq:schema} obey $d\ln u_m/dt=2(\alpha\lambda_m(\Cr)-\kappa-\varphi(U))$ with $U=\sum_nu_n$. This is a replicator dynamics on the shares, with per-schema fitness $\alpha\lambda_m$ \cite{touzel2026fitnesslandscapesocialnorms,Gintis2010} and the budget pressure $\varphi(U)$ of \cref{sec:model}, $\varphi\approx0$ while the channel is slack and $\varphi'>0$ as it fills. The total reinforcement of schema $m$ is $\g_m^{\!\top}\Cr\g_m=u_m\lambda_m$. The pressure acts equally on every axis, so it shifts the invasion threshold without biasing which schemas survive. Because the budget caps $U$ rather than conserving it, two states survive that a normalized replicator would eliminate: a sub-threshold state in which every schema decays, and a slack regime ($\Theta>1$, $U<U_{\mathrm{cap}}$) in which all $M$ schemas coexist at the per-schema ceiling.

\paragraph{The invasion threshold.} A schema becomes active where its replicator rate \cref{eq:schema} turns positive, $\alpha\lambda_m>\kappa+\varphi$, at an unstable invasion strength $u_m^\dagger$: a schema seeded below it decays to zero, one seeded above it grows. Invasion precedes ignition, since $u_m^\dagger$ sits below the bifurcation strength $1/\Lambda$ of \cref{eq:cascade} by the noise floor $\sidn^2/2\gamma$ measured against the decay rate (\smref{S6}). The strengths therefore arrive at the ignition threshold under their own dynamics rather than resting at it, and that arrival is the $u$-transient of \cref{sec:regimes}.

\begin{figure}[t!]
\centering
\includegraphics[width=\columnwidth]{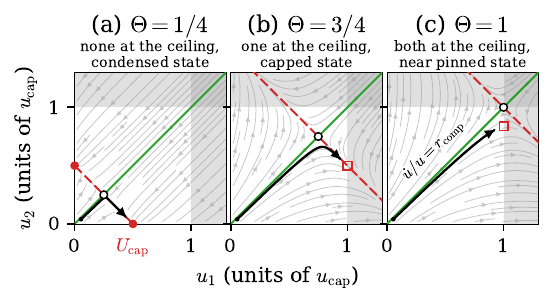}
\caption{\textbf{The pinning saddle across $\Theta$.} Phase portraits of the two-schema strength dynamics \cref{eq:schema} in $(u_1,u_2)$, units of $u_{\mathrm{cap}}$, at three budgets $U_{\mathrm{cap}}=\Theta Mu_{\mathrm{cap}}$ ($M=2$ throughout). The winner-take-all boundary then lies at $\Theta=1/M=1/2$, so the three panels sample one capacity band of \cref{fig:regimes} each. The pinned state (open circle saddle point), at $U_{\mathrm{cap}}/2$, moves out along the stable manifold $u_1=u_2$ (green) panel to panel as $U_{\mathrm{cap}}\to Mu_{\mathrm{cap}}$. The black curve is a trajectory from one seeded start (black dot, the same in all three panels), which grows along the stable manifold and then departs along the unstable one. The growth near the stable manifold is $\rcomp$ (\cref{eq:g}), labelled in (c). \textbf{(a)} $\Theta=1/4$: the budget alone halts growth well inside $u_{\mathrm{cap}}$, so the unstable manifold (red, condensation) runs to a genuinely reached bare condensate (filled circles). \textbf{(b)} $\Theta=3/4$: the ceiling is now reached first, arresting the same departure at $u_{\mathrm{cap}}$ (red square) before it reaches the excluded region $u_1{>}u_{\mathrm{cap}}$ or $u_2{>}u_{\mathrm{cap}}$ (shaded). \textbf{(c)} $\Theta=1$: capacity exactly meets demand ($M_{\mathrm{cap}}=M$), so the saddle sits at $u_1{=}u_2{=}u_{\mathrm{cap}}$ and the trajectory is arrested (red square) just short of it. Panel sub-titles name the resulting state.}
\label{fig:saddle}
\end{figure}

\paragraph{The pinning saddle.} With \cref{eq:lambda-m} of \cref{sec:fixedpoint} in hand, two competing results follow, drawn as phase portraits in \cref{fig:saddle}.
\begin{enumerate}
\item\emph{The pinning condition.} A schema is stationary only where its replicator rate \cref{eq:schema} vanishes, $\alpha\lambda_m=\kappa+\varphi^\ast$ with $\varphi^\ast=\varphi(U^\ast)$ the equilibrium budget pressure, so at an interior fixed point all survivors share one $\lambda_m=(\kappa+\varphi^\ast)/\alpha$. The shift $\varphi^\ast$ is common to every axis, so equal fitness still demands equal strength. $\lambda_m$ is strictly increasing in $u_m$: both factors in \cref{eq:lambda-m} grow with $u_m$ across the sub-critical range $\Lambda_m<1$, a property that extends to non-orthogonal repertoires (\smref{S5}). The interior fixed point therefore has a degenerate spectrum. This is the competitive-exclusion result for replicators. Monotonicity sharpens it from equal fitness to equal strength.
\item \emph{The condensation instability.} The pinned point is a saddle, and both of its directions follow from \cref{eq:schema} directly. Along the \emph{magnitude} direction $u_m=u+\delta$ (all schemas together), growth is arrested by whichever capacity ceiling is reached first (\cref{sec:regimes}): the per-schema cap $u_{\mathrm{cap}}$, or the total budget through $\varphi(U)$, which subtracts a restoring $2M\varphi'(U^\ast)\,\delta$ from every growth rate against the destabilizing $2\alpha\,(\partial\lambda/\partial u)\,\delta$. This stabilizes the direction once $M\varphi'(U^\ast)>\alpha\,\partial\lambda/\partial u$. Either way an imposed ceiling supplies the restoring force that fixes $U$; the bare dynamics do not. Along a \emph{composition} direction $\sum_m\delta u_m=0$ the total, and hence $\varphi$, are unchanged, and \cref{eq:schema} linearizes to $\dot{\delta u}_m\propto\delta\lambda_m\propto\delta u_m-\overline{\delta u}$: because $\lambda_m$ increases with $u_m$, any schema pushed above the mean grows, draining the rest---\emph{unstable}. One stable magnitude direction and $M-1$ unstable composition directions make the equal-strength state a saddle (\cref{fig:saddle}), and the dynamics leave it toward condensation. The departure is autocatalytic near a bifurcation, where $v_m$ (and hence $\partial\lambda_m/\partial u_m$) diverges. There the budget pressure arrests the runaway at a saturated cluster, and so regulates condensation. Whether it does so (a cluster) or does not (a bare condensate) is set by $\Theta=U_{\mathrm{cap}}/(Mu_{\mathrm{cap}})$, mapped globally in \cref{fig:regimes}.
\end{enumerate}

\paragraph{Classification of stationary states.} At a fixed point of
\cref{eq:schema} every schema is extinct ($u_m=0$, $\alpha\lambda_m<\kappa+\varphi^\ast$),
marginal ($0<u_m<u_{\mathrm{cap}}$, $\alpha\lambda_m=\kappa+\varphi^\ast$, at the invasion
threshold), or saturated ($u_m=u_{\mathrm{cap}}$). Since pinning forces equal
fitness to equal strength, no stationary state has a graded active
distribution, so there are exactly two fixed-point families. (a)~The
\emph{degenerate} state: all survivors equal, stabilized by the ceiling into a cluster of $\min(M,M_{\mathrm{cap}})$ saturated schemas. Its flat spectrum bifurcates as a single avalanche (class~i). (b)~The \emph{spiked} state: one
condensate holding the budget with a fringe pinned at the invasion threshold,
reached when the budget cannot cap even one schema (class~v). $\Theta$ exchanges them---the degenerate cluster shrinks from $M$ schemas
($\Theta\ge1$) to $M_{\mathrm{cap}}$ ($1/M\le\Theta<1$) to one ($\Theta<1/M$, the
spike).

\paragraph{Leaving the saddle.} The dynamics leave the saddle one axis at a time, along the engine chain \textit{schema $\to$ differentiation $\to$ signal $\to$ coordination $\to$ resources}. An inactive axis sits with its covariance eigenvalue at the noise floor $\sidn^2/2\gamma$. When an axis activates, it generates resources that reweight $\Cr$ toward the agents who differentiated along it. Because their $\omega$-identities also spread in other directions, this inflates the residual variance off the active axis, and the next schema grows until its eigenvalue clears the floor and ignites. Successive axes leave the floor in turn, and this sequence is the endogenous cascade.

\subsection{The regime map and its exact boundaries}\label{sec:regimes}

\begin{figure*}[t!]
\centering
\includegraphics[width=\textwidth]{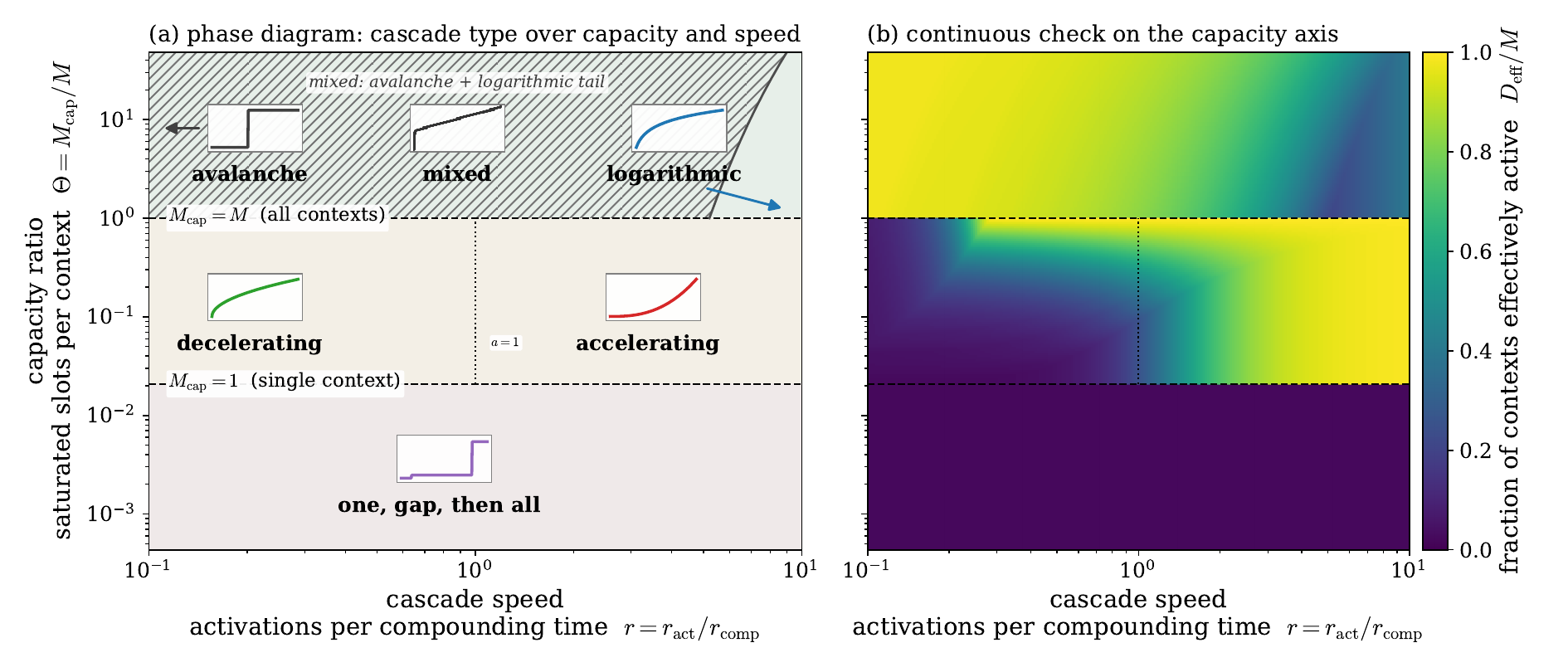}
\caption{\textbf{Cascade types and repertoire size over replicator dynamics.} Endogenous cascade regimes over the two coordinates that separate them. The horizontal axis (\emph{cascade speed}) is $r=\ract/\rcomp$, the number of axes that activate per compounding time $1/\rcomp$, fixed by the dispersion of the seeded repertoire through \cref{eq:nu}. The vertical axis (\emph{capacity ratio}) is $\Theta=(U_{\mathrm{cap}}/M)/u_{\mathrm{cap}}=M_{\mathrm{cap}}/M$, the budget's share per schema in units of the per-schema ceiling, equivalently the saturated slots per context ($M_{\mathrm{cap}}=U_{\mathrm{cap}}/u_{\mathrm{cap}}$). \textbf{(a)} \textit{Phase diagram}: where each cascade type sits in the plane. The capacity coordinate alone splits it into three bands with sharp counting-threshold boundaries---\emph{slack}, \emph{budget-limited}, and \emph{winner-take-all}---the contexts the society \emph{can} sustain being all $M$, a graded cluster of $M_{\mathrm{cap}}$, or one. Inset sketches show the sequence of activations the $u$-transient passes through, set by $r$: avalanche to logarithmic in the slack band ($\Theta\ge1$), decelerating ($a>1$) to accelerating ($a<1$) across $a=1$ in the budget-limited band, and one onset, a gap, then all the rest in the winner-take-all band. The left-hand arrow marks $r\to0$, where the cascade is a pure avalanche. The random Mar\v{c}enko--Pastur class (iv of \cref{fig:classes}) is absent because the loop selects endpoints, and an unstructured random spectrum is an initial condition rather than an endpoint (\cref{sec:regimes}). Within the slack band a saturated cluster coexists with a geometric tail (hatched) below the exact boundary $r=M/L(\Theta)$. The right-hand arrow runs from the logarithmic sketch past that boundary to a purely geometric spectrum, and the inset shows $N(\beta)$ at a coexistence point---an early jump as the cluster ignites at once, then the tail's gradual rise. \textbf{(b)} \textit{Continuous check on the capacity axis}: the contexts \emph{effectively active}, the participation ratio $D_{\mathrm{eff}}=(\sum_k\mu_k)^2/\sum_k\mu_k^2$ of the spectrum $\{\mu_k=u_k\}$, normalized by $M$, a continuous counterpart to the count $M_{\mathrm{cap}}/M$ on the vertical axis. It depends on how far the $u$-transient has progressed, hence on $r$, and falls below capacity wherever it is incomplete---bright on the accelerating side, darker on the decelerating side. The difference between the panels measures distance from the stationary endpoint. Exact boundaries, their mean-field status, the ceiling $u_{\mathrm{cap}}$, and the held-fixed groups ($\alpha\langle\lambda\rangle/\kappa$, $\beta\mu_1/\gamma$, $M/d$) in \cref{sec:regimes}.\label{fig:regimes}}
\end{figure*}

\paragraph{The $u$-transient and its two rates.} The candidate schemas do not cross threshold simultaneously: near threshold the $\lambda_m$ of the growing cohort are nearly equal, so the common growth rate converts the seeded spread in $u_m$ into a staggered sequence of threshold crossings. This staggered sequence is the $u$-transient. Two rates characterize it. The \emph{compounding rate} is read from \cref{eq:schema} for an active, pre-saturation schema (where the channel is slack and $\varphi\approx0$),
\begin{equation}\label{eq:g}
\rcomp\equiv 2(\alpha\lambda_\star-\kappa),
\end{equation}
$\lambda_\star$ the common value taken by the resource-weighted variance \cref{eq:lambda-m} across that cohort. $\rcomp$ is set by the gain-spectrum gap at the growing axis. \Cref{fig:saddle} shows this compounding as the growth along $u_1=u_2$, before the trajectory departs along the unstable direction. The \emph{activation rate} $\ract$ is the rate at which successive survivors cross threshold. \Cref{fig:classes}b counts the same crossings against the loop gain of an exogenous sweep. Here the spectrum rises at fixed $\beta$, so they are counted against time instead. The decay and the budget pressure are common to every schema, so they cancel from the log-differences, $d\ln(u_m/u_n)/dt=2\alpha(\lambda_m-\lambda_n)$. The common part translates the spectrum rigidly in $\ln\mu$ and moves every threshold together. Only the differences in $\lambda_m$ reshape it, so the log-spacings hold while the growing cohort's $\lambda_m$ stay near-equal. This is the quasistatic spectrum the classes of \cref{sec:cascade} assume, and $\lambda_m$ increasing with $u_m$ eventually breaks it. Candidates seeded as in \cref{sec:model} compound at the common rate $\rcomp$ and so reach a fixed threshold at times spread by $\Delta\ln u/\rcomp$, giving
\begin{equation}\label{eq:nu}
\ract\propto \rcomp/\sseed,
\end{equation}
The readout $r\equiv\ract/\rcomp\propto1/\sseed$ is therefore fixed by the dispersion of the candidate repertoire and is not set independently. Both $\ract$ and $\rcomp$ are emergent. $\sseed$ is a property of the repertoire, which for a deployed system is operator-held (\cref{app:control}).

The geometric and Zipf classes are morphologies of the $u$-transient toward those endpoints, not fixed points. A monitor watches roles form, so these morphologies are the observable object. Slow age-ordered relaxation is geometric (class~ii, $k_0=\ract/\rcomp$), share-proportional growth under a budget at capacity is Yule--Simon and hence Zipf (class~iii, $a\sim \rcomp/\ract$), and an unorganized exogenous repertoire is Marchenko--Pastur (class~iv). The replicator thus selects the cascade in two steps: $\Theta$ fixes the endpoint, and $r=\ract/\rcomp$ fixes its morphology and lifetime. No spectrum is imposed.

The cascade morphology depends on $\sim5$ dimensionless groups, but only two separate regimes: the capacity ratio $\Theta$ and the activations per compounding time $r$. The capacity $M_{\mathrm{cap}}$ is distinct from the realized survivor count $K$, since survivors need not be saturated. The two also have different ranges: $M_{\mathrm{cap}}$ exceeds $M$ in the slack regime ($\Theta>1$) and drops below $1$ in the winner-take-all regime ($\Theta<1/M$), whereas $K\le d$.

\paragraph{Two imposed ceilings.} The bare reinforcement supplies no turnover of its own, so no strength exists at which it saturates intrinsically (\smref{S6}). Two imposed ceilings arrest the growth of $u$ instead, and they are distinct: a per-schema cap $u_{\mathrm{cap}}$, and the total budget $U_{\mathrm{cap}}$ enforced by the soft pressure $\varphi(U)$ of \cref{eq:schema}. We take $u_{\mathrm{cap}}$ schema-independent. A schema at $u_{\mathrm{cap}}$ is \emph{saturated}, while the budget is \emph{slack} for $U<U_{\mathrm{cap}}$ and \emph{at capacity} as $U\to U_{\mathrm{cap}}$. Saturation throughout refers to the per-schema ceiling and never to the budget. Both are properties of the legible-status channel of \cref{eq:schema}: its finite power limits $U$, its finite range caps any single $u_m$. That independence leaves $\Theta$ free. A per-axis share of one budget, $u_{\mathrm{cap}}=U_{\mathrm{cap}}/M$, would fix $\Theta=1$ and remove the coordinate. The channel's third number is its noise scale $\sread$ (the measurement-quality limit of Ref.~\cite{Castro2026}), the finest status difference it resolves. The channel-to-strength mapping rests on a convention, so this fixes $u_{\mathrm{cap}}$ up to proportionality, which is all $\Theta$ requires.

\paragraph{Reading the map.} \Cref{fig:regimes} is the principal $(\Theta,r)$ slice, a phase diagram locating the cascade types (\cref{fig:regimes}a) with a continuous check on its count-based capacity axis (\cref{fig:regimes}b). The five spectra of \cref{fig:classes}, all at $R=50$, already span $D_{\mathrm{eff}}/M$ from $0.05$ (Zipf $a=2$) to $1$ (degenerate), so the realization panel reads out cascade class and not capacity alone. The two coordinates act on different objects. $\Theta$ selects the stationary endpoint: the saturation ceiling caps the condensation instability, in which all strength collects on one schema, when capacity exceeds demand ($\Theta>1$, giving a degenerate spectral cluster), and leaves that instability unchecked when the budget cannot sustain even one schema ($\Theta<1/M$, giving a spiked spectrum). Its boundaries are therefore sharp phase lines. $r$ measures how far apart in time successive candidates cross threshold. When the seeded repertoire is widely dispersed, the earliest contexts compound longest and incumbents entrench; when it is narrow, candidates ignite together and churn instead. The two limits are a smooth crossover between the geometric and Zipf classes. Here $r=\ract/\rcomp$ is the generative shape parameter fixed by the seed dispersion through \cref{eq:nu}. Four endogenous regimes arise. An early ceiling re-equalizes the winners into a \emph{saturated cluster} ($\Theta\ge1$), near-degenerate, class~(i). A standing cohort encodes age as strength, $u_m\propto e^{\rcomp(t-t_m)}$, giving the \emph{geometric} regime ($r\ll1$), class~(ii). A budget at capacity grows shares in proportion, a Yule--Simon process \cite{simon1955} giving \emph{Zipf} ($1/M\le\Theta<1$), class~(iii). Unregulated condensation leaves one winner over a fringe pinned at the invasion threshold, the \emph{spiked} regime ($\Theta<1/M$), class~(v). \Cref{fig:regimes}a places them in the plane. Three of the four boundaries are counting thresholds in $\Theta$ and one is a threshold in $r$, and all four are exact analytic curves on the principal slice (\cref{app:boundaries}). The map takes the repertoire near-orthogonal, parametrizes capacity by $\Theta$ alone, and describes the $u$-transient rather than a stationary hierarchy. \smref{S7} sets out what each of these assumes and where it fails.

\section{Monitoring: reading the cascade before it starts}\label{sec:monitoring}

\begin{figure*}[t!]
\centering
\includegraphics[width=\textwidth]{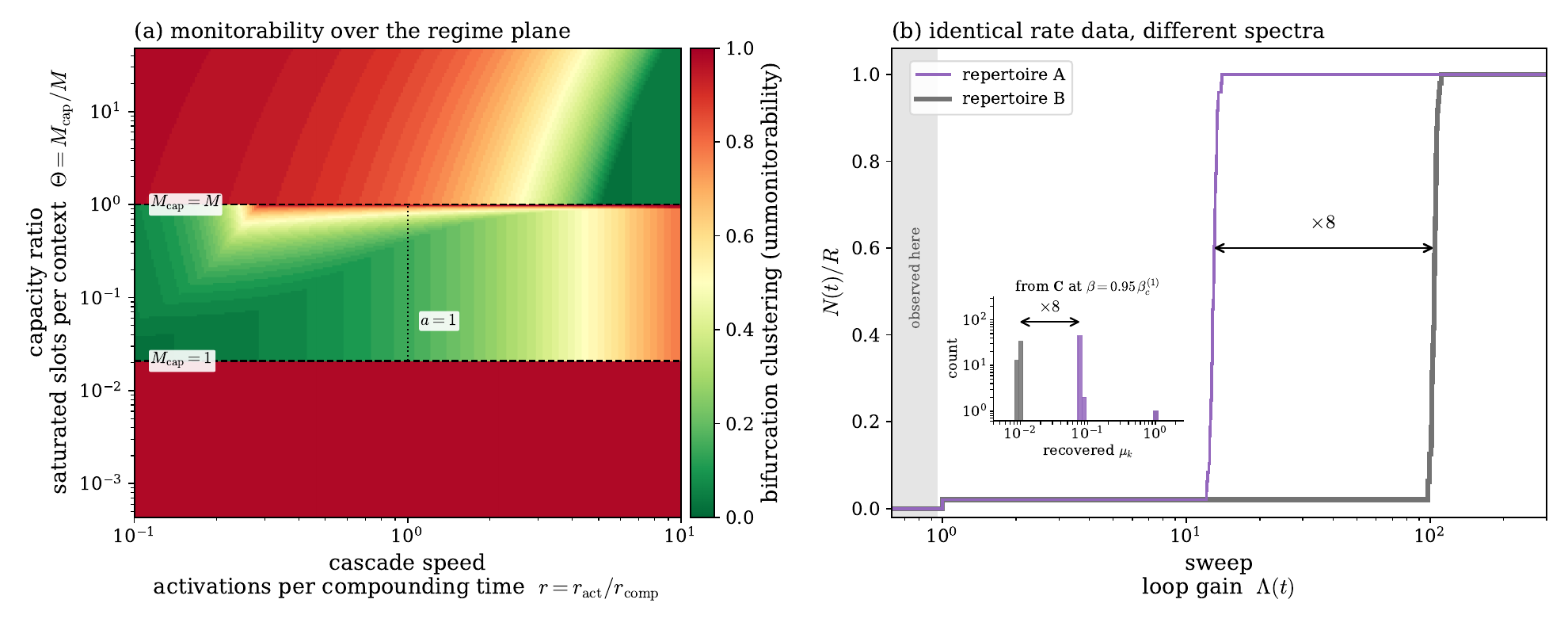}
\caption{\textbf{Monitoring the cascade.} \textbf{(a)} Monitorability over the $(\Theta,r)$ plane: the clustering of bifurcation onsets (colour; the maximal share arriving within $\Delta\ln\beta=0.15$, hence unmonitorability), red where onsets cluster into a single burst (avalanche, spiked) and green where they resolve into a sequence (logarithmic, decelerating). The $\Theta$ boundaries are sharp, the $r$ dependence a smooth crossover. Regime layout as in \cref{fig:regimes}. \textbf{(b)} The measurement term, under a linear sweep of the loop gain $\Lambda(t)$. Two repertoires whose counting functions coincide over the whole plotted range: each activates one axis at $\Lambda=1$ and then stays quiescent. In repertoire A the remaining $R-1$ modes sit just below threshold and avalanche at $\Lambda=13$; in repertoire B the same modes sit far below and do not activate until $\Lambda=104$. The two spectra differ only below $1/\Lambda$, which the bifurcation rate never samples, so rate extrapolation cannot separate them. Inset: the gain spectrum recovered from the subcritical covariance $\C=\tfrac{\sidn^2}{2\gamma}(\Id_{d\times d}-\Lam)^{-1}$ at $\Lambda=0.95$, before either has bifurcated. The two recovered bulks are separated by the same factor of $8$ that separates the two rises, since an onset sits at $1/\mu$. The paired arrows mark it. The inversion amplifies a $7\%$ difference in covariance eigenvalues into that factor, so the number of agents sampled sets the practical limit (\cref{sec:monitoring}).}
\label{fig:monitor}
\end{figure*}

\noindent Consider a monitor who samples the population's $\omega$-identities, which \cref{sec:model} makes available through a status ledger or a platform estimate (\smref{S4}). A ledger is read directly. An estimate the platform reconstructs from its encounter log carries the error of \cref{app:control}, which adds to the measurement term below. The monitor need not be the operator, but it needs the operator's access, since either route to the $\omega$-identities is platform-held. As the loop gain rises, role axes activate one at a time, and each activation appears as a split in the population along one such axis.
The monitor's task is to forecast: it needs a quantity, measurable before the next activation, that predicts the rest of the sequence, ideally with enough precision to determine the cascade class. The eigenvalues of the $\omega$-identity covariance $\C$ (\cref{eq:C-linear}) are such a quantity. A class is a shape of the gain spectrum $\rho$, so estimating $\rho$ is estimating the class. \Cref{eq:counting} makes $N$ depend on the products $\Lambda\mu_k$ alone, so mode $k$ ignites at $\ln\Lambda=-\ln\mu_k$. An error $\delta\ln\mu$ common to the spectrum therefore displaces every predicted onset by $-\delta\ln\mu$ in $\ln\Lambda$. Over a lead time $\Delta t$ the forecast is in error by
\begin{equation*}
\delta N\simeq\underbrace{\frac{dN}{d\ln\beta}}_{\substack{\text{bifurcation}\\[-1pt]\text{density}}}\bigl(\,\underbrace{\delta\ln\mu}_{\substack{\text{spectrum as}\\[-1pt]\text{measured}}}+\underbrace{\rcomp\,\Delta t}_{\substack{\text{drift over}\\[-1pt]\text{lead time}}}\,\bigr).
\end{equation*}
The two error terms are properties of the monitor's situation, and the density is a property of the repertoire.

\paragraph{The bifurcation density.} The density, the number of onsets per unit $\ln\beta$, is $dN/d\ln\Lambda=R\,\mu\rho(\mu)$ (\cref{eq:rate}). It converts a spectral error into a count error: where the spectrum is dense, a given $\delta\ln\mu$ blurs more onsets together, and the spacings that identify the class stop being resolvable. \Cref{fig:monitor}a maps it over the $(\Theta,r)$ plane as the largest share of onsets arriving within a fixed window in $\ln\beta$. An operator who reshapes the spectrum therefore changes how monitorable the system is, whatever the monitor does. Whether the loop reshapes it unaided is open. The resource weighting in $\Cr$ may generically widen the gaps between the $\mu_k$, and with them the window a monitor has (\smref{S8}).

\paragraph{The measurement term.} It depends on how $\{\Lambda_k\}$ is estimated, and there are two. The generic early-warning route is the relaxation time of the $k$-th mode of $\C$, $\tau_k=1/\gamma|1-\Lambda_k|$, which diverges at threshold. Estimating it means resolving that mode's autocorrelation in time, so the monitor needs a record spanning $O(\tau_k)$ at least, and longer for precision. The divergence is the critical slowing down whose rise is a generic early-warning signal \cite{scheffer2009}. If the engine parameter is swept at rate $\dot\Lambda$, the dimensionless ratio $\mathcal{R}=\tau_k\dot\Lambda$ controls feasibility. Only the local rate enters, so a sweep of any smooth shape is covered by its slope at the crossing, and an adoption curve is worst for a monitor where it is steepest. When the forcing is fast relative to the system's internal timescale, the sweep crosses the transition before the slowing can be measured. This is rate-induced tipping, where slowing-down indicators generically fail \cite{ashwin2012,ritchie2016}. Slow deployment ($\mathcal{R}\ll1$) preserves genuine early warning, and fast deployment renders it infeasible.

The other route reads the same modes at a single instant. Subcritically $\C$ is fixed by \cref{eq:C-linear}, so $v_k=\tfrac{\sidn^2}{2}\tau_k$ by the Ornstein--Uhlenbeck balance between the $\omega$-identity noise and the mode's relaxation rate: the $k$-th covariance eigenvalue is the $k$-th relaxation time. Inverting \cref{eq:vk} eigenvalue by eigenvalue, a measured subcritical spectrum $\{v_k\}$ returns the gain spectrum directly. That identification uses the flat noise floor: at finite encounter rate the shot noise of \cref{app:micro} adds a term proportional to $\mu_k$ to the numerator, so $v_k$ overstates $\tau_k$ on the strongest modes and biases the recovered $\Lambda_k$ upward there. A sample of agents at one instant estimates it with no time record, and the modes still to activate sit far below threshold, where $\tau_k\approx1/\gamma$ and the population has long since equilibrated. This route is limited by the number of agents rather than by the observation time. Where only actions are visible, the platform must instead run the parallel simulation of \cref{app:control}, inferring the projections $\w_i\cdot\hat\g_m$ from encounter-averaged actions through $\pibar$, which restores a dependence on observation time. That inference is hardest below threshold, where $P_e\to\tfrac12$ and a single encounter carries little about $\Delta_{im}$.

Sampling fluctuations give each measured eigenvalue a finite width, and eigenvalues closer together than that width cannot be separated. Inverting the empirical eigenvalue distribution of $\omega$-identity embeddings therefore recovers $\rho(\mu)$, and with it the full $N(\beta)$ (\cref{fig:classes}a), its bifurcation rate (\cref{fig:classes}b), and any avalanche masked by an early onset (\cref{fig:monitor}b). Rate extrapolation recovers none of this, since it samples $\rho$ only at $\mu\ge\gamma/\beta$ and must assume $\zeta$ below (\cref{eq:accel}). Before the first onset, and inside a quiescent interval, no spacings exist to fit, so only the inversion applies.

\paragraph{The drift term.} It grows with the lead time. The eigenvectors of $\Lam$ are the context directions, which the loop holds fixed, and its eigenvalues follow the schema strengths, which compound at $\rcomp$. A drift common to every mode shifts $\ln\mu$ uniformly, which displaces the cascade in $\beta$ and leaves its class unchanged. Only the spread of $\rcomp$ across surviving modes changes the class, so the class forecast outlives the timing forecast. Both hold over $\Delta t\lesssim1/\rcomp$, long whenever the schemas drift slowly compared with the sweep, since $\rcomp$ is of order $\kappa$. Where the repertoire is not near-orthogonal the eigenvectors move with the strengths as well, and that limits the forecast instead (\cref{sec:regimes}).

\paragraph{Classifying from the spacings.} The classification converts a low observed rate into a decision procedure. The first few inter-bifurcation spacings identify the class: constant ratios $\beta_{k+1}/\beta_k$ indicate geometric (estimate $k_0$, predict the rest); ratios growing as $((k{+}1)/k)^a$ indicate Zipf (the exponent's position relative to $a=1$ decides decay versus explosion); a rate rising as $(\Lambda-1)^{1/2}$ indicates an imminent random bulk. A long quiescent interval is not evidence that the cascade has ended, since it is also the spiked signature. Rate data alone cannot distinguish the two, and the subcritical spectral inversion resolves the ambiguity. The complementary design levers on the cascade class are collected in \cref{app:control}.

\paragraph{Designing the class.} The class structure also makes the cascade designable, but the prescription is not monotone in $r=\ract/\rcomp$: it reverses between the capacity bands, so the seeding that resolves the cascade in one band clusters it in the other. Those levers also frame a conjecture about deployment, in which parameters $\Theta$ and $r$ (fixed here) might evolve systematically, moving the system toward the avalanche-prone corner of \cref{fig:monitor}: more mature systems sitting at lower $\Theta$ (more coordination competing for finite capacity) and, as a platform's task taxonomy fills in around a settled core, at higher $r$ (a narrower spread of context weights). Engineering improvements to coordination might even erode monitorability. The operator's incentive runs toward coordination efficiency, which is what the monitoring levers of \cref{app:control} cost. What makes monitoring possible at all is a by-product: the status ledger an operator maintains for coordination is also what exposes the $\omega$-identities a monitor reads (\smref{S4}). Because those levers are platform-owned and their cost in coordination efficiency falls on the operator, while the consequences of an undetected transition do not, the incentive structure of platform design is itself a safety variable. Engineering practice outside the present theory will thus impact how these parameters vary across a system's lifetime. Making that drift an endogenous trajectory would require promoting $\Theta$ and $r$ to slow dynamical variables, a two-timescale extension we leave open.

% ======================================================================

\section{Discussion}\label{sec:discussion}

\noindent The model realizes Sewell's duality as a fixed-point equation---schema $\leftrightarrow$ context direction, resource $\leftrightarrow$ accumulated coordination payoff, structure $\leftrightarrow$ resource-weighted covariance, agency $\leftrightarrow$ deviation from prescribed role, transposability $\leftrightarrow$ context overlap $\hat\g_m\cdot\hat\g_n$. The spectrum of the social loop gain matrix then organizes onset, cascade shape, monitorability, and design.

\paragraph{Distributional AGI: a control surface.} Recent AI safety work \cite{tomasev2026} hypothesizes that general-level AI capability may emerge not in one model but from the coordination of many sub-general agents. Our framework makes this quantitative: the thick phase is the coordinated collective that realizes the (reward-optimal) correlated equilibrium. Shared substrates---task queues, codebases, finite compute---make same-action encounters interfere, which gives the payoffs an anti-coordination structure. Complementary roles (planner/executor, coder/reviewer) then extract gains no member could achieve alone, sustained by self-generated signals rather than assigned by an orchestrator. There is growing evidence of the benefits of coordination for tasks assigned to groups of agents \cite{anthropic2026}. There is also a growing number of examples where deployed agent collectives, left to self-organize when solving complex tasks, spontaneously seek out and use social technology such as message boards to coordinate (e.g. crowd-sourcing solution strategies and delegating subtasks) \cite{openai2026incidentreport,vonarx2026collusion}. For open-ended, ongoing tasks in which stable roles confer advantages, our theory assumes such behaviour implements the signalling needed to stabilize roles.

Several consequences for the distributional AGI hypothesis follow from identifying its coordinated collective with the thick phase. The first two concern what the transition is, the last two what an operator can do about it. First, an operator can cross the transition by improving the coordination layer alone---routing, protocols, channel noise ($\rhopair$, $\sread$)---with no change in any agent, so per-agent capability evaluations will not detect it. Where the repertoire is degenerate, every axis ignites together, so even small changes in the control can drive the onset of system-wide coordination (class~i of \cref{fig:classes}d).

Second, orchestration cannot compensate below the cognitive floor. Agent populations with $\etaag<\sread\gamma/4\beta$ cannot ignite the thick phase, even for strong orchestration. This is analogous to pattern storage and retreival in neural networks. Each schema must hold its own basin in the $\omega$-identity landscape, and \cref{eq:agent-mf} descends that landscape with the context vectors as its stored patterns, so the repertoire a population can carry is the capacity of an associative memory. Sharper inference deepens those basins, giving $M_{\mathrm{eff}}\le d^{\,n-1}$ with $n$ the policy sharpness \cite{krotov2016}, which rises with capability \cite{gershman2015}. More informally, crude agents sustain a coder and a reviewer, not an organization. But, the ceiling is exponential in $n$: one more unit of policy sharpness multiplies it by $d$ (\smeq{S12}). Capacity is a long-studied object for such networks \cite{hopfield1982,amit1985}, and \smref{S3} sets out where the correspondence holds and where roles depart from pattern retrieval.

Third, stored identity may impact behaviour similar to persona steering. Character traits in a language model can be monitored and controlled along directions in its activation space \cite{chen2025personavectorsmonitoringcontrolling}. An agent's $\omega$-identity $\w_i$ accumulates its own past role-taking and re-enters its context-for-action through memory retrieval. Competing alongside other parts context, collective-controlled agent identity could in principle influence action, thereby closing the loop. The theory adds a stability criterion: below ignition an imposed separation relaxes back to the noise floor, and above it the population holds the separation with no steering at all, because the separation induces roles. The monitor of \cref{sec:monitoring} is its population-level counterpart: it projects $\omega$-identities onto role axes where persona work projects activations onto a trait direction.

Fourth, the factors that set ignition and shape the cascade are \emph{platform-owned} apart from the agent capacity $\etaag$, so an operator who fixes the environment before ignition holds a control surface (\cref{app:control}). It is a design choice among settings, not control-theoretic steering during takeoff, since a fast sweep crosses the transition before the slowing can be measured (\cref{sec:monitoring}).

\paragraph{Monitoring in practice.} The analysis also fixes what a monitor outside the platform can do. The subcritical covariance carries the whole gain spectrum, so a cascade is readable before any role appears. That reading requires three conditions: the $\omega$-identities have to reach the operator, through a ledger or an estimate (\smref{S4}), so access is granted rather than assumed; the sweep has to be slow enough for the modes to be resolved; and the onsets must not already be clustered, which is a property of the repertoire rather than of whomever watches it. \Cref{sec:monitoring} sets out each. The operator has no incentive to supply any of the three conditions.

\paragraph{Scope: transformation.} We scoped down onto the case where the strengths $u_m$ are endogenous, while the axes $\hat\g_m$ are not. The loop therefore goes beyond reproduction, since which schemas govern is itself an outcome. It stops short of transformation in Sewell's sense, which requires a schema to change what it is about. Moving a schema requires the angular sector and \cref{app:directions} sets out the decisions in this extension. Nor does the model capture Sewell's claim that transformation is perpetual: our dynamics sort into roles and settle to a pinned or condensed fixed point, whereas a Sewellian society never settles. Sustained schema turnover lies outside the closed loop and we leave it for this extension to articulate.

\paragraph{Other Extensions.} Two extensions remain within the emergence regime. The first is including consensus outcomes. The theory here uses anti-coordination games, in which two players gain by differing: complementary roles confers reward. However, in general coordination also arises from consensus, where agents gain by matching their behaviour instead. Existing informational active matter realizes that case, with a population mean for its order parameter \cite{vansaders2023,ziepke2022}. A repertoire holding both would couple the differentiation sector treated here to a consensus sector, with gain eigenvalues of both signs: positive for differentiation, negative for consensus. Role axes inflate the covariance, while consensus axes shift the mean (the covariance-spectrum monitor sees only the former). The second extension realizes coherent roles across contexts. The mean field averages over which of the two roles an agent takes, so both earn the same, while the game pays them $6$ and $2$. That averaging makes the coordination payoff insensitive to the sign of the status difference. This leaves an agent's role signs uncorrelated across contexts (\smref{S3}). Retaining the difference lets agents who repeatedly take the higher-paying role accumulate more. It is the mechanism within the present game that would convert differentiation into a status ordering across contexts.

\paragraph{Toward an information thermodynamics.} The bound constraint of \cref{sec:model} is stated in information units, as is the consensus bound it parallels \cite{vansaders2023}. Here, the reason is that no energy scale exists to converts bits. The maximal gain $w_0$ occupies the position $k_{\mathrm B}T$ holds in a Szilard bound, its value fixed by the payoff matrix rather than a bath. The mean-field reduction removes the second ingredient: averaging the action over the observation noise discards the measurement record, so no cost of holding that record can be set against the payoff. Carrying the accounting at the level of the individual encounter, where the measurement record still exists, would put the payoff and the cost of that record in the same units, and is the step measurement-induced order needs to become thermodynamic. Because the loop is autonomous rather than cyclic, the closest precedent is an autonomous measurement device whose record is never reset \cite{cocconi2024}, where an excess entropy production measures the cost of holding it. The role label and the written action map onto that construction in \smref{S9}, which also sets out where the correspondence breaks. Developing a self-consistent thermodynamic accounting of the latter and taking it
%The cyclic single-agent engines of Refs.~\cite{still2020,STILL} instead subtract the cost of a memory erased each cycle, and show that the strategy maximizing gain alone is not the one that maximizes net output.
from the single agent to the population is the open step \cite{parrondo2015}.

\paragraph{Building agent platforms as experimental systems.}

The introduction noted the lack of an experimental system to ground modeling choices arising when developing a theory of structuration. An LLM-driven social simulation is one. Its configurations are set rather than inferred, so the modeling decisions a quantitative theory must make can be settled by construction. Using a deliberately minimal formulation (one anti-coordination game, a Gaussian mean field, a loop that settles rather than perpetually transforms), the model presented here can serve in designing such systems. It is solvable, built around a single mechanism, and extended by perturbation in identified small parameters: the repertoire overlap, the fluctuation width $v/\sread^2$, and $1/\Nag$ (\smref{S6}, \smref{S7}). Four design criteria follow. Agents need persistent memory whose write rate and retention the platform sets. The task pool needs anti-coordination tasks and a seeded spread of traffic across them, which fixes $r$ (\cref{sec:regimes}). A partner's standing has to be legible at a resolution the platform controls, since $\sread$ and the matching policy $\rhopair$ cross the transition without touching any agent (\cref{app:control}). Encounters and memory embeddings both have to be logged, since the covariance $\C$ is what a monitor reads. The experiment is then a slow sweep of $\Lambda$, and an operator can locate a deployment relative to threshold beforehand, without running it supercritical (\smref{S2\,C}).
%Role differentiation that appears without the subcritical covariance signature would indicate a mechanism other than the one proposed here.

\paragraph{Fixing the assumptions empirically.} Such a simulation can settle modeling choices the theory now fixes by assumption. The $\omega$-identity noise is the clearest case. We take it isotropic and independent of the drive, which gives the cascade a common floor to be read against. The encounter shot noise instead aligns with the context directions and grows as $\beta^2/\nu$ (\cref{app:micro}). The ignition thresholds $\Lambda_k=1$ follow from the drift and hold under either choice. The flat floor, the subcritical inversion of \cref{sec:monitoring}, and the separation between invasion and ignition depend on it. The simulation logs the encounter statistics and the $\omega$-identity dispersion together, so which noise dominates is open to measurement. The capacity is the second case. We impose both ceilings, the per-schema range $u_{\mathrm{cap}}$ and the total budget $U_{\mathrm{cap}}$, and take them independent, since the bare reinforcement supplies neither (\smref{S6}). Their ratio $\Theta$ then selects the stationary endpoint and places a deployment in one of the three capacity bands. Both are properties of the status channel, so its power and its range fix $\Theta$ by measurement rather than by choice. Truthful presentation is the third. We assume it, and the anti-coordination payoffs make it plausible, since inflated display recruits deference until it collides with genuine high status (\cref{app:micro}). Establishing it as an equilibrium is a signaling question, requiring cost-of-signal machinery rather than the mean-field apparatus used here \cite{lachmann2001}. A simulation that logs presented status alongside realized play measures the gap. Two further assumptions can be settled the same way: the repertoire geometry $\{\hat\g_m\cdot\hat\g_n\}$, taken near-orthogonal, and the closed pool, to which no schema is added after seeding. The theory supplies the consequences that follow once such assumptions are fixed, and fixing them empirically is where a science of agent populations would begin. That science will also have to accommodate agents whose departures from rationality are not the human ones. Here bounded rationality enters only as read noise and policy sharpness, the forms it takes for people. Language-model agents depart from best response in ways with no such compact description, so theories built on normative agents, this one included, should not be expected to fit them without modification. The open question is which of those departures the collective dynamics is sensitive to. Representation raises the companion question. Here an agent's identity and a task context enter only as vectors in one space, compared through the projection $\w_i\cdot\g_m$. Persona work supports half of that, since traits in a language model are directions in its activation space \cite{chen2025personavectorsmonitoringcontrolling}. Task context provided to the agent as a block of text also admits vector embedding, though a more tractable representation for variation might be learned as a dedicated module. The simulation embeds agent memories and task descriptions together, so whether status tracks a projection, and whether the covariance of those embeddings carries the low-rank structure $\Lam$ requires, are open to measurement. Both the pseudo-rational and language-mediated aspects of LLM-driven agent simulation can be tested in controlled multi-agent game settings.

\paragraph{The method, beyond the model.} The applications and extensions above stay inside the model's content. Two features of the construction do not. First, the loop gain matrix is itself dynamical, so the spectrum that sets the transitions is an output of the dynamics rather than an input to it. Second, the subcritical covariance is the static response to that gain, so inverting it recovers the spectrum before any transition occurs. Both transfer to any population whose interaction enters as a matrix rather than a scalar, whatever game its agents play.

\begin{acknowledgments}
The author acknowledges conversations with John Bechhoefer and Milo\v{s} Bro\'{c}i\'{c},
and grant support from the Future of Life Institute and IVADO.
\end{acknowledgments}

\section*{Data availability}
\noindent The simulation and figure-generation code that supports the findings of this study is
openly available at \url{https://github.com/mptouzel/role_diff}. No
experimental datasets were generated or analyzed.

\appendix

\section{Microfoundations of the mean-field dynamics}\label{app:micro}

\noindent The main-text role dynamics is the mean-field drift of an underlying encounter process. We record the reductions that produce it, since each is a modeling assumption with its own scope.

\paragraph{Encounter clock and the write rate.} Agent $i$ plays at Poisson times with rate $\nu$. At an encounter in context $m$ against partner $j$ it commits a signed decision $a\in\{-1,+1\}$ and inscribes a fixed increment $b_0\,a\,\g_m$ into $\w_i$, which otherwise relaxes:
\begin{equation}\label{eq:micro-sde}
d\w_i = -\gamma\,\w_i(1+|\w_i|^2)\,dt + b_0\sum_m \g_m\,dN_{im}(t) + \sidn\,d\bm B_i,
\end{equation}
with $N_{im}$ a counting process of rate $\nu f_m$ with signed jumps $a$. Taking the compensator, $\mathbb E[dN_{im}]=\nu f_m\langle a\rangle_m(\w_i)\,dt$, so the drift depends on $b_0$ and $\nu$ only through the product $\beta\equiv\nu b_0$: encounter frequency and per-encounter inscription are not separately identifiable at mean field. This replacement is the window average of \cref{sec:model} written as a differential. It discards a martingale with increments of order $\sqrt{\nu f_m\,dt}$, the shot noise quantified below. The same $\nu$ multiplies the coordination income (\cref{sec:fixedpoint}), so setting $\nu=1$ (time measured in average inter-encounter intervals) is a global units choice, not a per-equation one. $\beta/\gamma=\nu b_0/\gamma$ is then the number of times an $\omega$-identity is rewritten within its own decay horizon.

\paragraph{Order of the two limits.} The window average of \cref{sec:model} carries out the sum over partners on its own. At fixed $\Nag$ the feedback term is $\beta\sum_m f_m\,\g_m\sum_j p^{(m)}_{ij}\,\bar a\bigl((\w_i-\w_j)\cdot\g_m\bigr)$ with $\bar a$ the noise-averaged action of \cref{eq:micro-avg}, so the pairing probabilities already enter as weights. The dynamics is not closed, since its right-hand side carries the other $\Nag-1$ $\omega$-identities. The limit $\Nag\to\infty$ closes it. It removes the imprint of the focal agent on its own partners, which is $O(1/\Nag)$ because $\nu$ is a fixed per-agent budget, and the run-to-run spread of the sum over the realized $\omega$-identities. The pairing average therefore follows from dense encounters alone (\smref{S8}).

\paragraph{Signed decision and its noise average.} At the decision level the action is a hard sign of the integrated evidence, $a=\sign(\tfrac1T\sum_{t}s_t)$ on samples $s_t=\Delta+\sread\xi_t$ of the true status difference $\Delta=(\w_i-\w_j)\cdot\g_m$. What enters the drift is its expectation over the observation noise,
\begin{equation}\label{eq:micro-avg}
\bar a(\Delta)=\mathbb E_\xi\big[\sign(\bar s_T)\big]=\operatorname{erf}\!\Big(\frac{\sqrt T\,\Delta}{\sqrt2\,\sread}\Big),
\end{equation}
\begin{equation}
\bar a'(0)=\gain,\quad \pibar'(0)=\frac{\sqrt{T/2\pi}}{\sread}.
\end{equation}
The decision rule is a function of the observation $s$. The mean-field response is a function of the true status difference $\Delta$, the noise having been integrated out. Averaging $\bar a$ over the partner in turn gives $\langle a\rangle_m(\w_i)$, the response that enters \cref{eq:agent-mf}. The two symbols differ by that second average. Both are instances of one operation: smoothing an erf by a Gaussian returns an erf, with the two variances added in quadrature inside its argument (\smeq{S25}). Taken over the observation noise it produces \cref{eq:micro-avg}, and taken over the partner it softens the slope by the factor of \smref{S6}. It is this slope $\pibar'(0)$, not the hard sign, that sets the gain: an un-averaged $\sign$ would give a singular (delta-function) susceptibility and an ill-posed threshold. Integrating $T$ samples sharpens the response as $\sqrt T$ (the read memory), whereas the $\omega$-identity retention $1/\gamma$ is the write memory. The two are distinct stages of the cycle.

\paragraph{Context normalization.} One context is drawn per encounter, with frequencies normalized so that $\sum_m f_m=1$. The feedback term $\beta\sum_m f_m\langle a\rangle_m(\w_i)\g_m$ is then an expectation over which context is played rather than a sum that grows with $M$, and the gain matrix $A=\sum_m f_m\,\g_m\g_m^{\!\top}$ has $\operatorname{tr}A=\sum_m f_m u_m$, which stays $O(1)$ as contexts are added provided the capacity grows with the repertoire ($U_{\mathrm{cap}}\propto M$, i.e.\ fixed $\Theta$ of \cref{sec:regimes}) rather than being held fixed: adding contexts at fixed budget divides the same write rate more finely rather than driving $\omega$-identity harder. This read budget $\sum_m f_m=1$ is the counterpart on the read side of the capacity budget $\sum_m u_m\le U_{\mathrm{cap}}$ on the write side (\cref{sec:closedpool}). The dependence on how many contexts probe a $d$-dimensional space enters through the aspect ratio $M/d$ (\cref{sec:cascade}), not through an overall scale.

\paragraph{Pairing kernel and the well-mixed reduction.} Partners are drawn with probabilities $p_{ij}^{(m)}$ ($\sum_j p_{ij}^{(m)}=1$), collected in the pairing kernel $\bm{P}^{(m)}$. Linearizing the partner-averaged response near the symmetric fixed point,
\begin{equation}\label{eq:micro-lap}
\sum_j p_{ij}^{(m)}\,\bar a\big((\w_i-\w_j)\cdot\g_m\big)\approx \gain\,\g_m^{\!\top}\big((\Id_{\Nag\times\Nag}-\bm{P}^{(m)})\bm{\Omega}\big)_i,
\end{equation}
with $\bm{\Omega}$ the $\Nag$ $\omega$-identities stacked, so pairing enters the gain through the graph Laplacian $\bm{L}^{(m)}=\Id_{\Nag\times\Nag}-\bm{P}^{(m)}$ of the interaction network, and ignition is governed by the spectrum of $\sum_mf_m\,\g_m\g_m^{\!\top}\otimes \bm{L}^{(m)}$---a coupling of schema space to social space. For \emph{well-mixed} pairing $p_{ij}=1/\Nag$, $\bm{P}=\tfrac1{\Nag}\mathbf 1\mathbf 1^{\!\top}$ projects onto the population mean, $\bm{L}$ acts as the $\omega$-identity off that mean, and the partner-average reduces to coupling to $\langle\w\rangle$ (a Curie--Weiss field). This recovers $\Lam=\Lambda\,\bm{H}^{\!\top}\bm{H}$ and a scalar pairing factor: uniform $p_{ij}$ gives $\rhopair=0$ and factor $1$, while an annealed matcher that conditions on status gives the general $(1-\rhopair)$ of \smref{S2\,B} (the uniform diagonal reweighting of \smref{S2}). The results of this paper are the annealed case. Structured $p_{ij}$---homophily, communities, degree heterogeneity---makes $\bm{L}^{(m)}$ nontrivial. It shifts thresholds and lets network structure, not only $\|\g_m\|$, select which role axis ignites first. Because homophily makes $p_{ij}$ depend on $\{\w_i\}$, the interaction network then co-evolves with the repertoire. What we bracket is that quenched structure, not status-conditioned matching: an annealed matcher leaves no network to accumulate, so it decouples network from repertoire and closes the loop analytically.

\paragraph{Shot noise.} At finite $\nu$ the compensated jump stream contributes shot noise of intensity $\nu b_0^2\sum_mf_m\g_m\g_m^{\!\top}=\beta b_0\,\bm{H}^{\!\top}\mathrm{diag}(\mathbf{f})\bm{H}$. This is proportional to $\bm{H}^{\!\top}\bm{H}$ rather than isotropic, so it is aligned with the context directions and largest along the active axes. It is not a scalar renormalization of $\sidn$: retaining it replaces the flat floor $\sidn^2/2\gamma$ by a mode-dependent floor that grows with $\mu_k$. The poles of the covariance are unchanged, because the shot-noise covariance shares the eigenbasis of $(\Id_{d\times d}-\Lam)^{-1}$ and reweights only its numerator. The thresholds $\Lambda_k=1$ and the relaxation-time spectroscopy are therefore identical in the two cases, and only the pre-onset amplitudes differ. The intensity is $O(\beta b_0)=\beta^2/\nu$, so it grows with the drive rather than holding still as $\beta$ sweeps, and the two noise sources are not interchangeable: $\sidn$ supplies a fixed isotropic reference and this one neither. It vanishes in the dense-encounter limit ($\nu\to\infty$, $b_0\to0$ at fixed $\beta$), which leaves the isotropic $\sidn^2\Id_{d\times d}$ that gives the flat noise floor used in \cref{sec:cascade}. The same granularity carries the model's irreversibility: partners observe each other independently, so a miscoordinated encounter kicks both $\omega$-identities the same way, which no pairwise potential generates, and $P_e$ therefore sets the coordination payoff and the entropy production together (\smref{S9}). The floor is flat to the extent that $\beta b_0=\beta^2/\nu\ll\sidn^2$, that is where a single encounter moves an $\omega$-identity by much less than $\sidn^2/\beta$. We take this limit throughout.

\paragraph{Why the $\omega$-identity noise is kept.} Dropping $\sidn$ and letting the shot noise above be the only stochasticity would be more economical, since that noise is the measurement error the Szilard bound already concerns. Three properties decide against it here. The noise along an axis is then proportional to $\mu_k$, so it vanishes with the strength, $\lambda_m\to0$ as $u_m\to0$, and no schema can grow before role structure exists (\cref{sec:closedpool}). The floor becomes mode-dependent, so the thresholds no longer share a common reference. The floor also grows as $\beta^2/\nu$, so it rises while the monitor reads the sweep. All three raise $\lambda_m$ preferentially on the strongest axes, so retaining them would steepen the strength hierarchy and move the system toward the avalanche-prone classes. The isotropic choice is therefore the conservative one for \cref{sec:monitoring}.

\paragraph{Status microfoundation.} The signal $s_{im}=(\w_i-\w_j)\cdot\g_m+\sread\xi$ is agnostic between a noisy read of a status difference held in a ledger and exact self-knowledge minus a noisy read of the partner's self-stored status. The mean-field results are identical. We adopt the second reading---each agent stores its own status and presents a noisy version---so that no status ledger is presupposed. This keeps the emergence claim non-circular (a public ledger is itself an institution, analyzed as operator technology in \smref{S4}) and matches the feedback term $\beta a_{im}$, which inscribes the agent's own act. Presentation is assumed truthful. The anti-coordination payoffs penalize inflated display (feigned high status recruits deference until it collides with genuine high status at $(-1,-1)$), so honesty is a plausible equilibrium of a strategic-display extension we do not develop here.

\section{Platform control surface}\label{app:control}

\noindent The ignition and cascade factors are platform-owned apart from the agent capacity $\etaag$, so an operator holds a pre-ignition control surface over which takeoff scenario is reachable (\cref{fig:monitor}). The parameters an operator sets are the write rate $\beta$ and the retention $1/\gamma$ of the $\omega$-identity memory, the channel factors $\sread$ and $\rhopair$---the visibility of a partner's standing and the matching policy---the context play frequencies $\mathbf{f}$ and pairing efficiencies $\bm{\phi}$, and the repertoire itself: its directions $\{\hat\g_m\}$ and their seed dispersion $\sseed$. Everything else in the loop is an outcome. There are three kinds of control levers. \emph{Ignition} is monotone in the channel factors $\rhopair$ and $\sread$. Every agent must clear the cognitive floor, so raising channel noise or degrading matching shuts the engine down whatever the agents' quality. This lever acts entirely outside the agents. Disassembly does not: removing individual agents leaves the resource-weighted inertia in place and the order stands, much as it does when spins are removed from a ferromagnet.
\emph{Cascade class} is steerable directly. The effective gain reweights the context second-moment matrix by each context's play frequency times its pairing efficiency, $\bm{H}^{\!\top}\mathrm{diag}(\mathbf{f})\,\mathrm{diag}(\bm{\phi})\,\bm{H}$, so per-context rate limits (the frequencies $f_m$) and matching quality (the efficiencies $\phi_m$) are interchangeable levers. Optimizing that schedule splits a spiked spectrum into a geometric, extrapolable sequence (\smref{S2\,D}). In a representative repertoire it more than halves the log-spacing variance a monitor must resolve ($1.35\to0.58$) and costs roughly $40\%$ of the coordination efficiency. The same tradeoff recurs across the levers below: monitorability is obtained at the expense of coordination efficiency.
\emph{The $u$-transient} is set by the repertoire's seeding rather than by the ledger's memory: $r=\ract/\rcomp\propto1/\sseed$ (\cref{eq:nu}), so tiering task types by launch traffic, rather than opening them at equal weight, lowers $r$. Where the budget caps growth ($1/M\le\Theta<1$) this shifts the cascade off the avalanche side toward the extrapolable classes. In the slack band ($\Theta\ge1$) the preference reverses at $r=M/L(\Theta)$, and a narrower seeding resolves the cascade instead (\cref{sec:regimes}). The operator defines the task taxonomy, so $\sseed$ is platform-owned in the same sense as $\mathbf{f}$ and $\bm{\phi}$ and acts on the same object, the spectrum of the effective gain (\smref{S4\,A}).
Two tensions run through these levers. Legibility trades against lock-in: the widely dispersed repertoire that yields the extrapolable classes is also the one in which the earliest and most strongly seeded contexts compound longest into entrenched incumbency, so the third lever improves legibility and deepens lock-in at once. A benign setting must also be \emph{held} rather than reached. The closed loop evolves toward condensation without intervention (\cref{sec:closedpool}), so the system returns to the avalanche-prone classes once the intervention is withdrawn. Which levers an operator holds at all depends on where $\omega$-identity is kept. A platform-maintained status ledger gives the operator $\beta$, $\gamma$, and the matching correlation $\rhopair$ together. If $\omega$-identity memory migrates back to self-storage, those levers pass to the collective, and that migration is itself an early and observable warning that control over the knobs is being lost (\smref{S4}). The dependence also runs the other way. A platform that measures $\C$ recovers the role axes as its eigenvectors and the schema strengths from the subcritical inversion (\cref{sec:monitoring}). The drift in \cref{eq:agent} is a sum over realized actions, so a platform that logs encounters can integrate that equation in parallel and serve $s$ from its own estimate of $\w_i$, which tracks the $\omega$-identity up to the unobserved $\omega$-identity noise $\sidn\bm{\eta}_i$. \Cref{eq:resource} does not follow, since its driver is a payoff rather than an action, and \cref{eq:schema} inherits that, because the strengths follow the resource-weighted covariance. The $\omega$-identity apparatus is therefore platform-ownable up to that noise, and only the reward path stays with the agents.

\section{Schema directions and the vector form}\label{app:directions}

\noindent \Cref{eq:schema} evolves schema strengths on fixed axes. The general object is the vector reinforcement
\begin{equation}\label{eq:schema-vec}
\dot\g_m=-\kappa\,\g_m+\alpha\,\Cr\,\g_m ,
\end{equation}
which has the same two ingredients---covariance-driven and resource-weighted---but also lets a schema change which axis it prescribes. Writing $\g_m=\sqrt{u_m}\,\hat\g_m$ separates \cref{eq:schema-vec} exactly into
\begin{equation}\label{eq:polar-split}
\dot u_m = 2u_m\bigl(\alpha\lambda_m-\kappa\bigr),\qquad
\dot{\hat\g}_m = \alpha\bigl(\Id_{d\times d}-\hat\g_m\hat\g_m^{\!\top}\bigr)\Cr\,\hat\g_m ,
\end{equation}
with $\lambda_m=\hat\g_m^{\!\top}\Cr\hat\g_m$. The radial equation is \cref{eq:schema}. The angular equation is Oja's rule \cite{oja1982}: a power iteration on $\Cr$ constrained to the unit sphere.

\paragraph{Collapse of the angular sector.} The angular equation has no $m$-dependence beyond its initial condition, since every schema is driven by the same matrix $\Cr$. Its stable fixed point is the leading eigenvector of $\Cr$ for all $m$, so $\hat\g_m\cdot\hat\g_n\to1$ and $\bm{H}^{\!\top}\bm{H}$ becomes rank one. By \cref{eq:cascade} a rank-one repertoire has a single threshold, so the cascade reduces to one bifurcation and the classification of \cref{fig:classes} does not apply. This is the collapse identified for the unweighted rotation $\dot\g_m=-\kappa\g_m+\alpha\C\g_m$ (\smref{S1\,A}). Resource weighting changes which matrix is iterated but not the fact that one matrix drives every schema.

The collapse is also the opposite of the transformation the theory is meant to describe. Rotation toward the dominant axis of $\Cr$ aligns every schema with the structure that already exists, which reproduces the leading role dimension rather than extending a schema to a context it was not built for. Part of that mechanism is already present as the repertoire's coherence, the overlaps $\hat\g_m\cdot\hat\g_n$ of \cref{sec:model}. Between orthogonal contexts the coupling is competitive only, through the shared budget.

\paragraph{Repelling mechanisms.} Sustaining distinct directions requires a term that separates schemas. Two are available in this model. Neither is contained in \crefrange{eq:agent}{eq:schema}.

\emph{(i) Imposed lateral inhibition.} Subtracting each schema's overlap with the others,
\begin{equation}\label{eq:inhibition}
\dot{\hat\g}_m = \alpha\bigl(\Id_{d\times d}-\hat\g_m\hat\g_m^{\!\top}\bigr)\Cr\,\hat\g_m-\eta\sum_{n\neq m}(\hat\g_m\cdot\hat\g_n)\,\hat\g_n ,
\end{equation}
drives the repertoire onto distinct eigenvectors of $\Cr$ ordered by eigenvalue---the deflation of principal-subspace learning \cite{sanger1989}. It is effective but stipulated: the rate $\eta$ has no counterpart in the payoff structure, so the resulting repertoire is a property of the learning rule rather than of the game.

\emph{(ii) Marginal-return reinforcement.} The redundancy can instead be removed at its source. Coordination income sums over contexts, so two aligned schemas give $\Delta_{i1}=\Delta_{i2}$ and an agent collects $W$ twice for one binary role distinction, while the role information the pair supplies is $I$ rather than $2I$. Additive income therefore violates the Szilard bound of \cref{sec:model} whenever schemas are redundant. Charging each schema only its marginal contribution replaces $\lambda_m$ by
\begin{equation}\label{eq:marginal}
\lambda_m^{\mathrm{marg}}=\hat\g_m^{\!\top}\Bigl(\Id_{d\times d}-\sum_{n\neq m}\hat\g_n\hat\g_n^{\!\top}\Bigr)\Cr\,\hat\g_m ,
\end{equation}
so a schema is reinforced only by variance no other schema already organizes. Rotation then moves toward uncovered variance, which is transposability in Sewell's sense, and the repelling term is derived from the payoff rather than added to the dynamics. The cost is that \cref{eq:resource} and the $\lambda_m$ of \cref{sec:closedpool} must both be re-derived, since income is no longer additive across contexts.

We adopt neither mechanism here. The questions this paper asks are ones the strength sector answers on its own.

\section{Regime boundaries}\label{app:boundaries}

\noindent On the principal slice the boundaries are exact analytic curves:
\begin{align}
\text{budget at capacity:}&\quad \Theta=1, \label{eq:b-theta1}\\
\text{winner-take-all possible:}&\quad \Theta=1/M, \label{eq:b-thetaM}\\
\text{Zipf $a=1$:}&\quad r=1 \ \ (a=1/r), \label{eq:b-a1}\\
\text{sat.\ cluster$|$geometric:}&\quad r=\frac{M}{L(\Theta)}, \label{eq:b-e1e2}\\
\text{with}&\quad L(\Theta)=\ln\frac{u_{\mathrm{cap}}}{u_0}=\ln\frac{U_{\mathrm{cap}}}{\Theta Mu_0}. \nonumber
\end{align}
These curves are exact in the self-averaged ($\Nag\to\infty$) mean-field limit used throughout. Finite populations fluctuate around them, and \emph{which} schema condenses is a symmetry-breaking choice the class does not fix. The map is in this sense a prediction: the $u$-transient, its classes, and these boundaries follow from the near-threshold Gaussian covariance (\cref{sec:closedpool}) extended into the post-bifurcation thick phase, whose full-loop verification is left to the accompanying simulation program.
Boundary \cref{eq:b-e1e2} is where the number of saturated modes $n_{\mathrm{sat}}=\max(0,M-rL)$ reaches zero: below $r=M/L$ a saturated cluster of $n_{\mathrm{sat}}$ modes coexists with a geometric tail (a mixed avalanche-plus-log class); above it the cluster is gone. This transition lies within the slack band, where \cref{fig:regimes} marks it by the avalanche and logarithmic sketches rather than a drawn curve. The colour field of \cref{fig:monitor}a is the bifurcation clustering, defined as the maximal share of bifurcation points $b_k=1/\mu_k$ falling within a window $\Delta\ln\beta=0.15$. It is a continuous diagnostic and varies smoothly across these sharp analytic lines. It reaches its floor slightly before \cref{eq:b-e1e2}, because a few residual saturated modes still cluster just below $n_{\mathrm{sat}}=0$.

\bibliography{references}

\end{document}